\documentclass[12pt]{article}
\pdfoutput=1
\usepackage{jheppub}

\usepackage{graphicx}
\usepackage[normalem]{ulem}
\usepackage{amsmath}
\usepackage{amssymb}
\usepackage{amscd}
\usepackage{enumerate}
\usepackage{amsfonts}
\usepackage{epsfig}
\usepackage{mathtools}
\usepackage{yfonts}
\usepackage{dsfont}
\usepackage{bbold}
\usepackage{breqn}
\usepackage{mathtools}
\usepackage[utf8]{inputenc}
\usepackage[english]{babel}
\usepackage{amsthm}

\DeclareMathOperator{\e}{\epsilon}

\DeclareMathOperator{\Tr}{Tr}

\newcommand{\be}{\begin{equation}}
\newcommand{\ee}{\end{equation}}
\newcommand\eea{\end{eqnarray}}
\newcommand\bea{\begin{eqnarray}}

\newcommand{\bZ}{{\mathbb Z}}

\title{Enhanced corrections near holographic entanglement transitions: a chaotic case study}
 
\author[a]{Xi Dong}
\author[b]{and Huajia Wang}
\affiliation[a]{Department of Physics, University of California, Santa Barbara, CA 93106, USA}
\affiliation[b]{Kavli Institute for Theoretical Physics, University of California, Santa Barbara, CA 93106, USA}

\emailAdd{xidong@ucsb.edu}
\emailAdd{huajia@kitp.ucsb.edu}

\abstract{Recent work found an enhanced correction to the entanglement entropy of a subsystem in a chaotic energy eigenstate. The enhanced correction appears near a phase transition in the entanglement entropy that happens when the subsystem size is half of the entire system size. Here we study the appearance of such enhanced corrections holographically. We show explicitly how to find these corrections in the example of chaotic eigenstates by summing over contributions of all bulk saddle point solutions, including those that break the replica symmetry. With the help of an emergent rotational symmetry, the sum over all saddle points is written in terms of an effective action for cosmic branes. The resulting Renyi and entanglement entropies are then naturally organized in a basis of fixed-area states and can be evaluated directly, showing an enhanced correction near holographic entanglement transitions. We comment on several intriguing features of our tractable example and discuss the implications for finding a convincing derivation of the enhanced corrections in other, more general holographic examples.}

\begin{document}

\maketitle 
\section{Introduction}

Entanglement and Renyi entropies of subsystems are important quantities that encode key properties of a quantum system as a whole.  In suitable limits where the number of relevant degrees of freedom becomes large, entanglement and Renyi entropies become analytically tractable and have been studied in (at least) two classes of examples.  The first involves energy eigenstates in the thermodynamic limit of a chaotic system~\cite{Garrison,Tolya1,Lu2017}.  The second involves holographic states whose entanglement entropy in the large-$N$ limit is given by the Ryu-Takayanagi (RT) or Hubeny-Rangamani-Takayanagi (HRT) formula in terms of the area of a bulk extremal surface~\cite{Ryu2006,Ryu2006_2,Hubeny_2007}, and whose Renyi entropies are determined from similar areas of appropriate cosmic branes~\cite{Dong:2016}.\footnote{A hybrid of these two classes of examples combining the thermodynamic limit and large-$N$ limit was studied in~\cite{Dong:2019}.}

As the size of the subsystem is varied, entanglement and Renyi entropies can experience phase transitions in the limit of a large number of degrees of freedom, signifying major rearrangements of the entanglement structure. Such phase transitions can be understood as the exchange of dominance between two saddle point solutions.  In the first class of examples involving chaotic eigenstates, the phase transition happens when the volume of the subsystem reaches half of the volume $V$ of the entire system.  Interestingly, a universal correction to the entanglement entropy that scales like $\sqrt{V}$ was found near the phase transition point~\cite{Vidmar:2017pak,Srednicki:2019}.  This correction is parametrically larger than the expected corrections away from the transition point, and we will refer to it as the enhanced correction near the entanglement transition.

The primary goal of this paper is to study the appearance of such enhanced corrections in holographic theories near entanglement transitions where two competing HRT surfaces have about the same area.  In the holographic limit $N \to \infty$ or ($G_N \to 0$ with $G_N$ being the gravitational constant), we expect the enhanced corrections to scale like $G_N^{-1/2}$ in general.  To find such corrections, one would in principle need to first calculate Renyi entropies
\be\label{sndef}
S_n = \frac{1}{1-n} \log \Tr \rho^n
\ee
at integer $n \geq 2$ holographically by summing over contributions of all bulk saddle points satisfying replicated boundary conditions, and then analytically continue to $n=1$ carefully to find the entanglement or von Neumann entropy.  The reason that we need to sum over all saddle points, instead of only keeping the most dominant one, is because there could be effects that scale like $e^{-(n-1)^2/G_N}$, resulting from summing over subdominant, exponentially suppressed saddle points at integer $n \geq 2$ but becoming non-negligible in the $n\to 1$ limit for the entanglement entropy.  As we will see, enhanced corrections near entanglement transitions have precisely this kind of behavior in the Renyi index $n$.

Practically speaking, the aforementioned task of summing over all bulk saddle points and then analytically continuing is generally very difficult to accomplish.  One important reason is that in principle we need to include replica non-symmetric saddle points -- bulk solutions that break the $\bZ_n$ replica symmetry on the boundary.  The number of replica non-symmetric saddle points depends on $n$, and therefore we cannot hope to simply continue each one of them analytically; we need to sum over them and then analytically continue, which is difficult in general.

Despite this general difficulty, we find that in the example of chaotic eigenstates, the corresponding holographic calculation of summing over all saddle points can directly be carried out in the thermodynamic limit.  This is a hybrid of the two classes of examples mentioned earlier, combining the thermodynamic limit and the holographic limit.  In this case, the calculation is possible due to an approximate rotational symmetry that emerges in the thermodynamic limit.  All bulk saddle point solutions -- including replica-nonsymmetric ones -- enjoy this emergent rotational symmetry.  We will therefore be able to sum over all bulk saddle points explicitly and express the resulting Renyi entropies in terms of an effective action for cosmic branes.  The full gravitational path integral for the Renyi entropies is then naturally organized in a basis of fixed-area states introduced in~\cite{Dong:flat,Akers:2018fow,Dong:2019piw}, with a final integral over the area.  The integral can be analytically continued to $n=1$ and the resulting entanglement entropy shows an enhanced correction near the entanglement transition, in precise agreement with Ref.~\cite{Srednicki:2019}.\footnote{Another holographic example where an analogous enhanced correction was recently discussed is the 2d gravity model in Ref.~\cite{replica_wormhole_west}.}

As we will see, our tractable example seems to teach us some ``lessons'' which we briefly summarize here and will discuss in more detail in later sections.  The ``honest'' calculation involves summing over all saddle points in the gravitational path integral.  This is difficult, but we can rewrite the result by postponing until the very end the integrals over the areas of the two competing HRT surfaces.  The other integrals are therefore calculated in fixed-area states and given by summing over saddle points that satisfy fixed-area constraints -- which we call fixed-area saddle points.  In our tractable example, we can explicitly sum over all fixed-area saddle points due to the emergent rotational symmetry and find the enhanced correction, but we can also get correct result (with subleading errors of order $1$) by simply keeping only the most dominant fixed-area saddle point (for each fixed area)\footnote{This is not in contradiction with our earlier statement below Eq.~\eqref{sndef} that we need to sum over all saddle points and not just the most dominant one, because here we are discussing \textit{fixed-area} saddle points before the final integral over the areas is to be performed.  It turns out that this final integral is the one that does not have a controlled saddle point approximation near entanglement transitions (when analytically continued to $n=1$), and by performing it directly one finds the enhanced correction.}.  The dominant fixed-area saddle point is replica-symmetric.  There are two replica-symmetric saddle points that are relevant near an entanglement transition (corresponding to two competing HRT surfaces), and which one dominates depends on the fixed areas.  A surprise worth pointing out is the following.  We might try to ``improve'' the calculation by summing over both replica-symmetric fixed-area saddle points (instead of handpicking the dominant one), letting the path integral decide which one dominates ``dynamically''.  We might hope that this could be a step towards summing over all saddle points and should be no worse than handpicking the dominant one.  However, this ``improved'' calculation does not work and in fact would not give an enhanced correction.  This failure may appear puzzling, but we should remember that the honest calculation involves summing over all saddle points, and there is no particular reason why summing over a subset -- consisting of the replica-symmetric saddle points -- should be better than keeping only the most dominant one.

Even though the enhanced corrections were first found in the von Neumann entropy, we expect them to appear also in Renyi entropies when the Renyi index $n$ is sufficiently close to $1$.  We find this explicitly in chaotic eigenstates without using holography in Section~\ref{sec:general}, after we briefly review the structure of these states and then calculate the Renyi entropies carefully by summing over all disorder Wick contractions (with help from Appendix~\ref{app:contraction})\footnote{This is the analogue of summing over all saddle points, whereas the calculation in Ref.~\cite{Srednicki:2019} is analogous to keeping only the most dominant fixed-area saddle point.}.

In Section~\ref{sec:holography}, we turn to the holographic calculation for chaotic eigenstates.  We first formulate the boundary value problem and find the relevant bulk saddle point solutions.  With the help of the emergent rotational symmetry, we explicitly sum over the saddle points and calculate the result in the fixed-area basis.  We conclude in Section~\ref{sec:disc} with discussions on the interpretation and outlook of our results, including the very interesting open question of how to find a convincing derivation of the enhanced corrections in other, more general holographic examples.

\bigskip

Closely related work has been done independently by Don Marolf, Shannon Wang, and Zhencheng Wang.  We have arranged with these authors to coordinate submission of our papers.

\section{Chaotic high energy eigenstates}\label{sec:general}
In this section, we begin by studying the enhanced correction to the entanglement of subsystem $A$ in chaotic high energy eigenstates, as in \cite{Srednicki:2019}. Instead of looking at the von Neumann entropy $S^A_{\text{ent}}$, we focus on the manifestation of such enhanced corrections directly in the Renyi entropy $S^A_n$. The proposal for the chaotic high energy eigenstates takes the form : 
\be\label{eq:random_z}
|E\rangle =  \sum_{E-\Delta<E_i+E_j<E+\Delta} c_{iJ}|E_i\rangle_A |E_J\rangle_{\bar{A}} 
\ee 
where $|E_i\rangle_A$ and $|E_J\rangle_{\bar{A}}$ are eigenstates of the subsystem Hamiltonians $\lbrace H_A, H_{\bar{A}}\rbrace$ with eigenvalues $\lbrace E_i, E_J\rbrace$ respectively. The coefficients $c_{iJ}$ are un-correlated random variable with unit variance. 
\be\label{eq:inde_Gaussian}
\overline{c_{iJ}c_{i'J'}}=\delta_{ii'}\delta_{JJ'}
\ee
where the over-line represent averaging over the randomness. One can also replace the constraint $E-\Delta<E_i+E_j<E+\Delta$  by a smooth Gaussian prefactor, and in order to make (\ref{eq:random_z}) an eigenstate of the total system, we should introduce an interaction term to the total Hamiltonian as in \cite{Srednicki:2019}:
\be 
H = H_A + H_{\bar{A}} + H_{\text{int}}
\ee
such that 
\be
\langle E | H_{\text{int}}^2 | E\rangle \propto \Delta^2 
\ee
In this work we are mainly interested in enhanced corrections that occur near the entanglement transition. They are not qualitatively affected by finite $\Delta$. In the following, we will work in the $\Delta \to 0$ limit and neglect any subtleties associated with finite $\Delta$. It is easy to work out the un-normalized reduced density matrix: 
\be
\rho_A = \sum_{E_i-2\Delta <E_j<E_i+2\Delta}\;\; \left(\sum_{E-E_i-\Delta< E_J<E-E_i+\Delta}c_{iJ}c_{jJ}\right) |E_i\rangle_A \langle E_j|_A 
\ee
This is ``stripe'' diagonal with width $2\Delta$. Upon averaging we have:
\be
\overline{\rho}_A = \sum_{E_i}\bar{d}(E-E_i)|E_i\rangle_A\langle E_i|_A 
\ee
and the normalization factor for $\bar{\rho}_A$ is given by: 
\be
\mathcal{N}=\sum_id(E_i) \bar{d}(E-E_i)
\ee
where $d\left(\mathcal{E}\right)$ and $\bar{d}(\mathcal{E}')$ are the number of states $|E\rangle_A$ and $|E'\rangle_{\bar{A}}$ with $\mathcal{E}-\Delta <E<\mathcal{E}+\Delta$ and $\mathcal{E}'-\Delta <E'<\mathcal{E}'+\Delta$ respectively. Similarly the $n$-th power of reduced matrix is equal to: 
\bea
\rho_A^n &=& \sum_{E_{i_1}}\left(\sum_{i_2,...,i_{n+1};J_1,...,J_n} \prod^{m=n}_{m=1} c_{i_mJ_m}c_{i_{m+1}J_m}\right)|E_{i_1}\rangle_A \langle E_{i_{n+1}}|_A
\nonumber\\
\sum_{i_2,...,i_{n+1};J_1,...,J_n} &=&...\sum_{E-E_{i_m}-\Delta<E_{J_m}<E-E_{i_m}+\Delta}\;\; \sum_{E_{i_m}-2\Delta <E_{i_{m+1}}<E_{i_m}+2\Delta}...
\eea
This is also a stripe diagonal matrix, albeit with width $2n\Delta$. The broadening should not be significant in the limit $\Delta\to 0$ while keeping $n$ finite. 

To compute the averaged n-th power of the reduced density matrix, one can neglect the difference between $\overline{\ln{\left[\text{tr}\rho^n_A\right/\left(\text{tr}\rho_A\right)^n]}}$ and $\ln{\left[\text{tr}\overline{\left(\rho^n_A\right)}/\left(\text{tr}\overline{\rho}_A\right)^n\right]}$, and compute: 
\be
\text{tr}\overline{\rho^n_A}= \sum_i \left(\overline{\sum_{a_1,...,a_n;b_1,...,b_n} c_{a_1,b_1}c_{a_2,b_1}c_{a_2,b_2}c_{a_3,b_2},...,c_{a_n,b_n}c_{a_1,b_n}}\right)
\ee
Using (\ref{eq:inde_Gaussian}), it receives contributions from all choices of Wick contractions between the random variables $c_{iJ}$. We work out the details of summing over dominant Wick contractions in Appendix~\ref{app:contraction}. One can understand different choices of Wick contraction as producing different saddles in the microstate calculation. This is done explicitly and one obtains that: 
\bea\label{eq:renyi_wick}
\text{tr}\overline{\rho_A^n} &=& \sum_i d(E_i)\bar{d}(E-E_i)\; G_n (E_i)\nonumber\\
G_n (E_i) &=& \begin{cases}  \overline{d}(E-E_i)^{n-1}\; \mbox{$_2$F$_1$}\left[1-n,-n;\,2;\,\frac{d(E_i)}{\overline{d}(E-E_i)}\right], \;d(E_i) < \overline{d}(E-E_i) \\ d(E_i)^{n-1}\; \mbox{$_2$F$_1$}\left[1-n,-n;\,2;\,\frac{\overline{d}(E-E_i)}{d(E_i)}\right], \;d(E_i) > \overline{d}(E-E_i)
\end{cases}
\eea
In the thermodynamic limit, we can replace the sum over states by integral over density of states: 
\bea\label{eq:trace_intergal}
\text{tr}\overline{\rho_A^n} &\propto & \int d\mathcal{E}\; e^{S_A(\mathcal{E})+ S_{\bar{A}}\left(E-\mathcal{E}\right)}\;
G_n(\mathcal{E})\nonumber\\
\!\!\!\!\! G_n(\mathcal{E}) &=& \begin{cases} e^{(n-1) S_{\bar{A}}\left(E-\mathcal{E}\right)}\; \mbox{$_2$F$_1$}\left[1-n,-n;\,2;\,e^{S_A(\mathcal{E})-S_{\bar{A}}\left(E-\mathcal{E}\right)}\right], \;S_A(\mathcal{E}) < S_{\bar{A}}\left(E-\mathcal{E}\right) \\e^{(n-1) S_A(\mathcal{E})}\; \mbox{$_2$F$_1$}\left[1-n,-n;\,2;\,e^{S_{\bar{A}}\left(E-\mathcal{E}\right)-S_A(\mathcal{E})}\right], \;S_A(\mathcal{E}) > S_{\bar{A}}\left(E-\mathcal{E}\right)
\end{cases}
\eea 
Therefore we can write the averaged microstate\footnote{From now on we shall refer to quantities of this type simply as ``microstate'' without specifying the averaged nature explicitly.} Renyi entropy $\overline{S_n}$ and the Renyi entropy $S^{\text{MC}}_n$ for the global micro-canonical ensemble $\rho= \sum_{E-\Delta< E' < E+\Delta}|E'\rangle \langle E'|$ as
\bea\label{eq:renyi_integral}
\overline{S_n}&=&\frac{1}{1-n}\ln{\left\lbrace \mathcal{N}^{-n} \int d\mathcal{E}\; e^{S_A(\mathcal{E})+ S_{\bar{A}}\left(E-\mathcal{E}\right)}\;
G_n(\mathcal{E})\right\rbrace}\nonumber\\
S^{\text{MC}}_n&=&\frac{1}{1-n}\ln{\left\lbrace \mathcal{N}^{-n} \int d\mathcal{E}\; e^{S_A(\mathcal{E})+n S_{\bar{A}}\left(E-\mathcal{E}\right)}\right\rbrace},\;\;\;
\mathcal{N}= \int d\mathcal{E}\; e^{S_A(\mathcal{E})+S_{\bar{A}}\left(E-\mathcal{E}\right)}
\eea

Based on the assumption of ergodicity, we take the following ansatz for the subsystem entropies: 
\be\label{eq:ergo_density}
S_A\left(\mathcal{E}\right)=f V s\left(\frac{\mathcal{E}}{Vf}\right),\; S_{\bar{A}}\left(E-\mathcal{E}\right)=(1-f) V s\left(\frac{E-\mathcal{E}}{V(1-f)}\right)
\ee
where $V$ is the total system size and $f=V_A/V$ is the subsystem fraction, furthermore $s(e)$ is the entropy density of the system as a function of the energy density $e$. From this we can write Eq. (\ref{eq:renyi_integral}) in terms of two exponent functions $F_1(x), F_2(x)$ as: 
\bea\label{eq:renyi_diff}
&&\overline{S_n}-S^{\text{MC}}_n = \frac{1}{1-n}\ln{\left\lbrace \frac{\int d\mathcal{E}\; \exp{\left[F_1(\mathcal{E})\right]}}{\int d\mathcal{E}\; \exp{\left[F_2(\mathcal{E})\right]}} \right\rbrace}\nonumber\\
&& F_1(\mathcal{E})= f V s\left(\frac{\mathcal{E}}{Vf}\right)+(1-f) V s\left(\frac{E-\mathcal{E}}{V(1-f)}\right)+\ln{G_n(f,\mathcal{E})}\nonumber\\
&& F_2(\mathcal{E})= f V s\left(\frac{\mathcal{E}}{Vf}\right)+n(1-f) V s\left(\frac{E-\mathcal{E}}{V(1-f)}\right)
\eea
where $G_n\left(f,\mathcal{E}\right)$ simply means replacing $S_A(\mathcal{E})$ and $S_{\bar{A}}\left(E-\mathcal{E}\right)$ in $G_n(\mathcal{E})$ using the ansatz Eq.~(\ref{eq:ergo_density}). In generic cases, the exponent $F_1(x)$ has two local maximum $(x^a,x^b)$ satisfying: 
\be
x^a<\text{min}(f,1-f)E<\text{max}(f,1-f)E<x^b
\ee
while $F_2(x)$ has only one. The saddle point equations take the form: 
\bea\label{eq:saddle_eq}
s'\left(\frac{\mathcal{E}_1}{Vf}\right)&=& s'\left(\frac{E-\mathcal{E}_1}{V(1-f)}\right)-\frac{\partial_{\mathcal{E}}G_n(f,\mathcal{E}_1)}{G_n(f,\mathcal{E}_1)}\nonumber\\
s'\left(\frac{\mathcal{E}_2}{Vf}\right)&=& ns'\left(\frac{E-\mathcal{E}_2}{V(1-f)}\right)
\eea
where $\mathcal{E}_1$ and $\mathcal{E}_2$ are the saddle points for $F_1$ and $F_2$ respectively.

In the remaining of the section we will compute (\ref{eq:renyi_diff}) using the saddle point approximation and consider the fluctuations. At the transition point $f=1/2$, we shall also see the transition in the nature of the saddle point calculation as $n$ is tuned towards $n=1$. 

\subsection{Away from $f = 1/2$}
Before focusing on the vicinity of $f=1/2$, we first check what happens away from it. When the subsystem is smaller than half of total system size by a non-infinitesimal amount: $f<1/2$. For generic (non-stringy) system, one expects that $s(x)\propto x^\alpha, 0<\alpha<1$, so $s'(x)$ is a monotonous decreasing function, so for $n>1$ we have:
\be\label{eq:E2_bound}
\mathcal{E}_2 < E f 
\ee

Since $s(x)$ is monotonously increasing, from Eq. (\ref{eq:E2_bound}) we can deduce that for $f<1/2$, we have:
\be 
S_A(\mathcal{E}_2)=f V s\left(\frac{\mathcal{E}_2}{f V}\right)<(1-f)V s\left(\frac{E-\mathcal{E}_2}{(1-f)V}\right)=S_{\bar{A}}\left(E-\mathcal{E}_2\right)
\ee 

As mentioned before, $F_1(\mathcal{E})$ has two local  maximum $\mathcal{E}^a_1<f E < (1-f)E< \mathcal{E}^b_1$. In this case the smaller one $\mathcal{E}^a_1$ is the dominant saddle, around which $G_n$ is given by the first line branch of Eq.~(\ref{eq:trace_intergal}). By assuming that  $S_A(\mathcal{E}^a_1)<S_{\bar{A}}\left(E-\mathcal{E}^a_1\right)$, we can approximate the hypergeometric function in $G_n$ as: 
\be 
\mbox{$_2$F$_1$}\left[1-n,-n;\,2;\,x\right] \approx 1+\frac{(n-1)n}{2}x,\;\; x\ll 1
\ee 
and the saddle point equation can then be approximated by
\be
 \left(1+\frac{n(n-1)}{2}\zeta_1 \right) s'\left(\frac{\mathcal{E}^a_1}{Vf}\right) \approx \left(n-\frac{n(n-1)}{2}\zeta_1 \right) s'\left(\frac{E-\mathcal{E}^a_1}{V(1-f)}\right)\\
\ee 
where
\be
\zeta_1 = \frac{e^{S_A\left(\mathcal{E}^a_1\right)}}{e^{S_{\bar{A}}\left(E-\mathcal{E}^a_1\right)}}.
\ee
This differs from the saddle point equation for $\mathcal{E}_2$ in (\ref{eq:saddle_eq}) only by terms of order $\zeta_1$, i.e. exponentially suppressed. So we can replace $\zeta_1$ by $\zeta_2 =  \frac{e^{S_A\left(\mathcal{E}_2\right)}}{e^{S_{\bar{A}}\left(E-\mathcal{E}_2\right)}}$, and expand the solution $\mathcal{E}^a_1$ around $\mathcal{E}_2$: 
\be\label{eq:soln_pert} 
\mathcal{E}_1 \approx \mathcal{E}_2 + \frac{\frac{V}{2}(n^2-1)f(1-f)s'\left(\frac{\mathcal{E}_2}{Vf}\right)}{f s''\left(\frac{E-\mathcal{E}_2}{V(1-f)}\right)-(1-f)s''\left(\frac{\mathcal{E}_2}{Vf}\right)} \zeta_2
\ee 
Therefore for $f<1/2$, $\mathcal{E}^a_1$ is close to $\mathcal{E}_2$ up to exponentially small parameter $\zeta_2 \ll 1$. This is also self-consistent with the assumption that $S_A(\mathcal{E}^a_1)<S_{\bar{A}}\left(E-\mathcal{E}^a_1\right)$. The difference between microstate and micro-canonical ensemble Renyi entropy can thus be computed via the saddle point approximation by: 
\be
\overline{S_n}-S^{\text{MC}}_n \approx \frac{F_1(\mathcal{E}_1)-F_2\left(\mathcal{E}_2\right)+\frac{1}{2}\ln{\left[F_2''(\mathcal{E}_2)/F_1''(\mathcal{E}_1)\right]}}{1-n} \propto \mathcal{O}\left(\zeta_2\right)\sim e^{-\sharp V}
\ee
This shows that for smaller-than-half subsystems, the microstate and micro-canonical ensemble agree in terms of Renyi entropy up to corrections suppressed exponentially.

When the subsystem becomes larger than half of the total system size $f>1/2$, the microstate result is no longer close to that of the micro-canonical ensemble.  Technically, the global maximum of $F_1(x)$ becomes the larger one $x_2$ of the two local maximum $x_1<x_2$. Unlike $x_1$, which is exponentially close to the global maximum of $F_2(x)$, $x_2$ is not related to that of $F_2(x)$. As a result, the saddle point approximations for $\int d\mathcal{E}\;e^{F_1(\mathcal{E})}$ and $\int d\mathcal{E}\;e^{F_2(\mathcal{E})}$ will thus yield deviations that is of finite fraction. Since both terms scale with the total volume, the deviation is also volume-law: 
\be
\overline{S_n}-S^{\text{MC}}_n \propto \mathcal{O}(V),\; f>1/2 
\ee

\subsection{Transition point $f = 1/2$}
We now focus on what happens at, or very close to, the transition point $f=1/2$. To clarify the details of calculation, we decompose $F_1$ into two parts: 
\bea 
F_1(\mathcal{E})&=& F_{\text{dom}}\left(\mathcal{E}\right)+ F_\Delta\left(\mathcal{E}\right)\nonumber\\
F_{\text{dom}}\left(\mathcal{E}\right) &=& S_A\left(\mathcal{E}\right)+S_{\bar{A}}\left(E-\mathcal{E}\right) + (n-1)\text{max}\left\lbrace S_A\left(\mathcal{E}\right),S_{\bar{A}}\left(E-\mathcal{E}\right)\right\rbrace\nonumber\\
F_\Delta \left(\mathcal{E}\right) &=& \ln{\left(\mbox{$_2$F$_1$}\left[1-n,-n;\,2;\,e^{-|S_A\left(\mathcal{E}\right)-S_{\bar{A}}\left(E-\mathcal{E}\right)|}\right]\right)}
\eea
We can interpret $F_{\text{dom}}$ as the contribution from choosing the dominant Wick contraction\footnote{This ansatz of picking only the dominant term was proposed in Ref.~\cite{Srednicki:2019}; see Eq.~(11) there.}, while $F_\Delta $ encodes the corrections from the subdominant Wick contractions. As mentioned before, we can think of different Wick contractions as corresponding to different ``saddles'' in the full calculation. This becomes more apparent in the context of AdS/CFT (see Sec.~\ref{sec:holography}). However, we should distinguish them from with the stationary-points $\mathcal{E}_{1,2}$ when evaluating (\ref{eq:renyi_integral}) in this section. 

At $f=1/2$, the exponent $F_1(\mathcal{E})$, as well as its decomposition $F_{\text{dom}}\left(\mathcal{E}\right)$ and $F_\Delta \left(\mathcal{E}\right)$, have a reflection symmetry: $\mathcal{E}\leftrightarrow E-\mathcal{E}$, so the saddles appear in pairs $\left(\mathcal{E}_1, E-\mathcal{E}_1\right)$ and make equal contributions, and we can write: 
\be 
\overline{S_n} = \frac{1}{1-n}\ln{\left\lbrace \mathcal{N}^{-n}\; 2\times \int^{E/2}_0 d\mathcal{E}\; e^{F_{\text{dom}}\left(\mathcal{E}\right)+ F_\Delta \left(\mathcal{E}\right) } \right\rbrace}
\ee

At the transition point, as we are mainly interested in the enhanced correction proportional to $\sqrt{V}$ in $\overline{S_n} - S^{\text{MC}}_n$, it would be helpful to strip away factors that only cause $\mathcal{O}(1)$ corrections to simplify computations. Let us define: 
\be
S^{\text{dom}}_n = \frac{1}{1-n}\ln{\left\lbrace \mathcal{N}^{-n}\; 2\times \int^{E/2}_0 d\mathcal{E}\; e^{F_{\text{dom}}\left(\mathcal{E}\right) } \right\rbrace} 
\ee
and observe that striped-out factor $e^{F_\Delta} =  \mbox{ $_2$F$_1$}\left[1-n,-n;\,2;\,x\right], x = e^{-|S_A - S_{\bar{A}}|} $ is bounded within $\mathcal{O}(1)$ factors for $0<x<1$: 
\be
1 \leq  e^{F_\Delta\left(\mathcal{E}\right)} \leq C_n,\;\;
 C_n = -\frac{\pi \text{Csc}{(2\pi n)}}{\Gamma(-2n)\Gamma(1+n)\Gamma(2+n)} = 1+ \frac{n-1}{2} + \mathcal{O}(n-1)^2
\ee
Therefore we have that: 
\be 
 \int^{E/2}_0 d\mathcal{E}\; e^{F_{\text{dom}}\left(\mathcal{E}\right)}  \leq \int^{E/2}_0 d\mathcal{E}\; e^{F_{\text{dom}}\left(\mathcal{E}\right) + F_\Delta\left(\mathcal{E}\right)} \leq \int^{E/2}_0 d\mathcal{E}\; e^{F_{\text{dom}}\left(\mathcal{E}\right)} C_n
\ee
We can therefore conclude that:
\be
S^{\text{dom}}_n - \frac{1}{n-1}\ln{C_n}  \leq \overline{S_n} \leq S^{\text{dom}}_n 
\ee
The bound is independent of $V$ for all $n$, and in the $n\to 1$ limit it becomes: 
\be
S^{\text{dom}}_n - \frac{1}{2} + \mathcal{O}(n-1) \leq \overline{S_n} \leq S^{\text{dom}}_n 
\ee
At this step, we can already see that the potential $\mathcal{O}\left(\sqrt{V}\right)$ enhanced correction for $\overline{S_n}-S^{\text{MC}}_n$ should be entirely captured by the dominant-term ansatz: 
\be
\overline{S_n}-S^{\text{MC}}_n = S^{\text{dom}}_n-S^{\text{MC}}_n + \mathcal{O}(1)
\ee
So in the following, we shall instead focus on $S^{\text{dom}}_n$, it is a good proxy for the microstate Renyi entropy $\overline{S_n}$. 

For generic $n$ away from $n=1$, the saddle point $\mathcal{E}_2$ for $F_2$ in (\ref{eq:saddle_eq}) is smaller than $E/2$: 
\be
0<\mathcal{E}_2 < \frac{E}{2}  
\ee
Furthermore, $F_{\text{dom}}\left(\mathcal{E}\right) \equiv F_2\left(\mathcal{E}\right)$ for $0<\mathcal{E}<E/2$, which fully encloses the saddle point $\mathcal{E}_2$, so we have that:
\be 
\int^{E/2}_0 d\mathcal{E}\; e^{F_{\text{dom}}\left(\mathcal{E}\right)} \approx \int^E_0 d\mathcal{E}\; e^{F_2\left(\mathcal{E}\right)}
\ee
up to exponentially suppressed corrections. As a result we will get that:
\be\label{eq:transition_finite_n} 
S^{\text{dom}}_n-S^{\text{MC}}_n \approx \frac{\ln{2}}{1-n}
\ee
which is $\mathcal{O}(1)$ and not enhanced. 

However, there is a caveat for this analysis near $n=1$. We have implicitly assumed that the contributions to $S^{\text{dom}}_n$ from the saddle point $\mathcal{E}_2$ can be approximated by twice of the standard full Gaussian integral: 
\bea\label{eq:Gaussian}
\int^{E/2}_0  d\mathcal{E}\; e^{F_{\text{dom}} (\mathcal{E})}\approx   e^{F_2(\mathcal{E}_2)} \int^{\infty}_{-\infty} dx\;\exp{\left(-\frac{1}{2}F_2''(\mathcal{E}_2)x^2\right)}= e^{F_2(\mathcal{E}_2)}\sqrt{\frac{2\pi}{F_2''(\mathcal{E}_2)}}
\eea

This is valid when the range of the integration around $\mathcal{E}\approx \mathcal{E}_2$ is much larger than the variance of the Gaussian: $ 1/\sqrt{\|F_2''(\mathcal{E}_2)\|}$, i.e. the saddle $\mathcal{E}_2$ is well separated from $E/2$. On the other hand, we know that $\mathcal{E}_2\to E/2$ as $n\to 1$. In other words, the two symmetric saddles $\left(\mathcal{E}_2, E-\mathcal{E}_2\right)$ of $F_{\text{dom}}$ in the full range $0<\mathcal{E}<E$ collide as $n\to 1$. When this happens, (\ref{eq:Gaussian}) is no longer a good approximation, i.e. dominant contributions to $S^{\text{dom}}_n$ do not consist of two independent full Gaussian integrals. As a result, (\ref{eq:transition_finite_n}) breaks down. 

\begin{figure}[h!]
\centering
\includegraphics[scale=0.2]{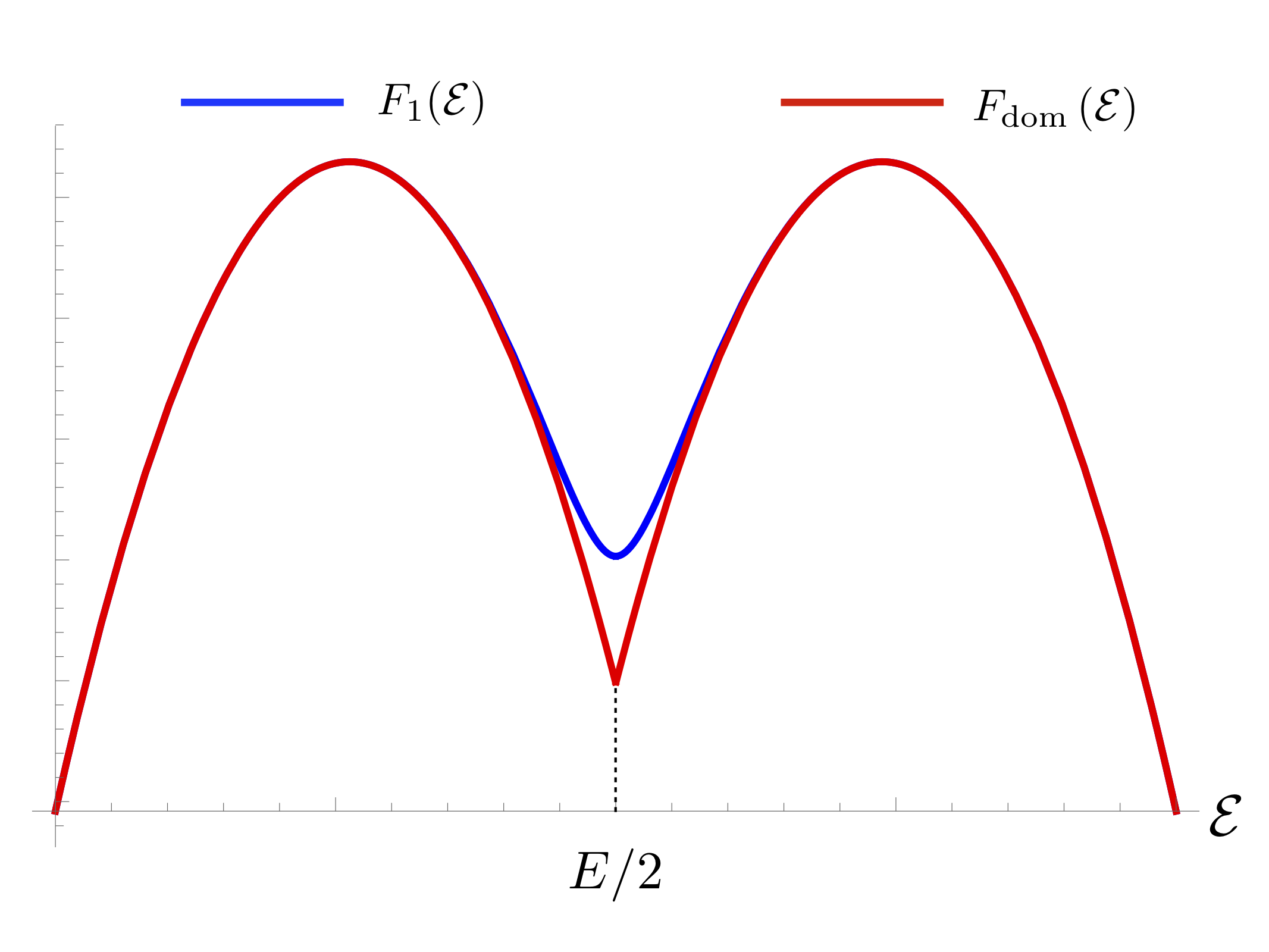}  
\includegraphics[scale=0.2]{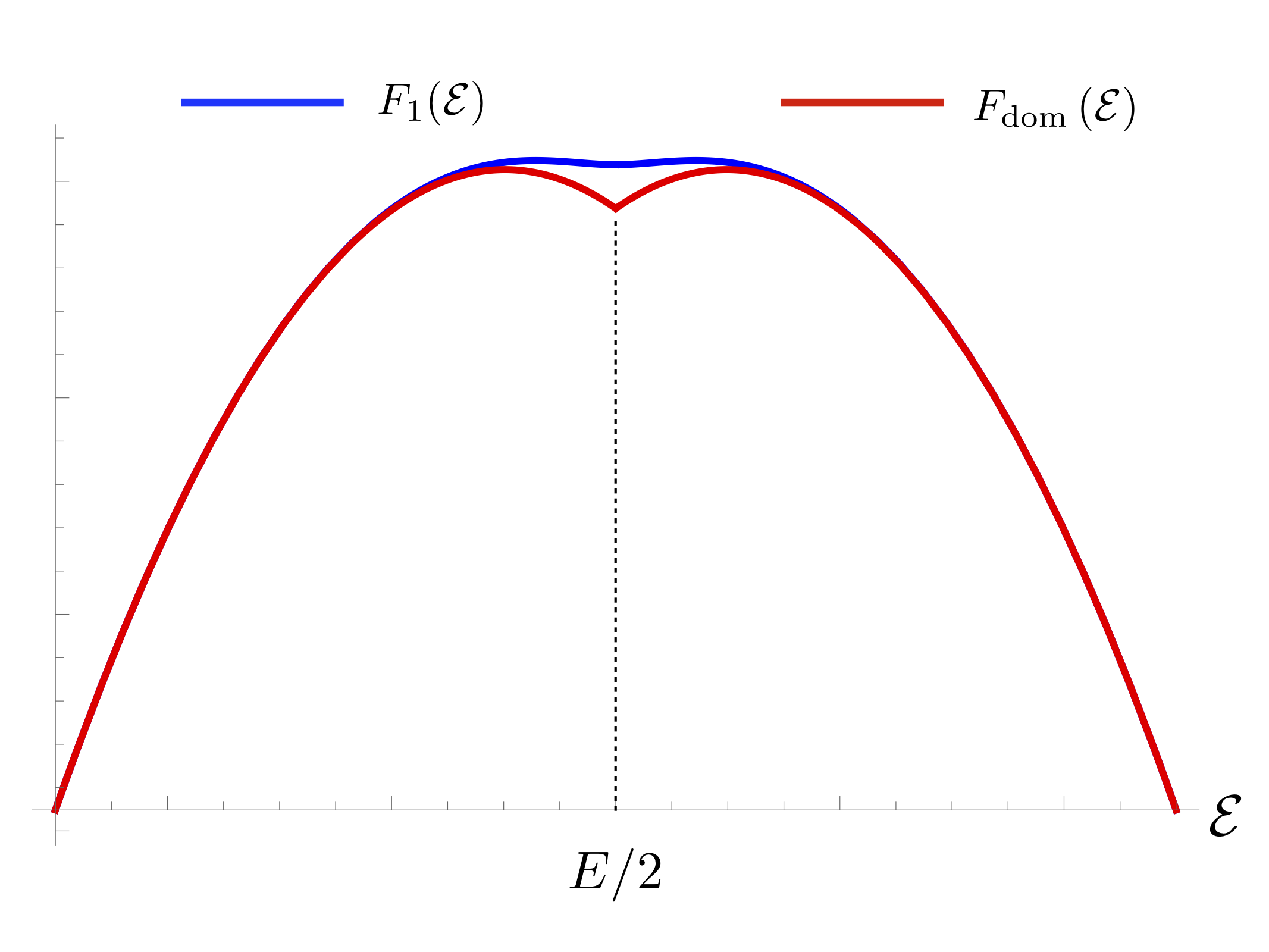}  
\caption{$F_1\left(\mathcal{E}\right)$ and $F_{\text{dom}}\left(\mathcal{E}\right)$ for different regimes of Renyi index $n$.  Left: $n-1\gg 1/\sqrt{C_V}$, the saddle points are well separated, there is no enhanced correction; Right: $n-1\sim 1/\sqrt{C_V}$, the saddle points are close within the curvature scale of $F_1$, an $\mathcal{O}\left(\sqrt{V}\right)$ enhanced correction appears, and an approximate flat interval for $F_1\left(\mathcal{E}\right)$ emerges between the saddles. 
}
\label{fig:F_plots}
\end{figure}

Let us expand around this limit in terms of $\delta = (n-1)$. The saddle point equation for $F_2(x)$ is easily solved to leading order by: 
\be
\mathcal{E}_2 = \frac{E}{2}+\frac{V}{4}\frac{s'\left(E/V\right)}{s''\left(E/V\right)}\delta + \mathcal{O}(\delta^2)
\ee 
Notice that in general we expect $s''(E/V)<0$. As a result the integration range for $S^{\text{dom}}_n$ around $\mathcal{E}_2$ has the positive side cut-off at order $\frac{V}{4}\frac{s'(E/V)}{s''(E/v)}\delta$ (see Figure~\ref{fig:F_plots}). Therefore to the leading order in $\delta$ the full Gaussian integral approximation (\ref{eq:Gaussian}) is valid if: 
\be\label{eq:enhancement_regime}
\frac{V}{4}\frac{s'(E/V)}{s''(E/V)}\delta \gg \frac{1}{\sqrt{F_2''(\mathcal{E}_2)}}\;\;\rightarrow\;\; \delta \gg \frac{2}{\sqrt{V}}\frac{\sqrt{| s''(E/V) |}}{s'(E/V)}
\ee

Below this range, we should instead proceeds as: 
\bea 
&&2\int^{E/2}_0 d\mathcal{E}\;\exp{\left[F_{\text{dom}}(\mathcal{E})\right]} \approx 2\int^{\frac{V}{4}\frac{s'(E/V)}{| s''(E/V) |}\delta }_{-\infty} d\mathcal{E}\;\exp{\left[F_2(\mathcal{E}_2)+\frac{1}{2}F_2''(\mathcal{E}_2)\mathcal{E}^2\right]}\nonumber\\
&=& \int^{\infty }_{-\infty} d\mathcal{E}\;\exp{\left[F_2(\mathcal{E}_2)+\frac{1}{2}F_2''(\mathcal{E}_2)\mathcal{E}^2\right]}+\int^{\frac{V}{4}\frac{s'(E/V)}{| s''(E/V)|}\delta }_{-\frac{V}{4}\frac{s'(E/V)}{|s''(E/V)|}\delta} d\mathcal{E}\;\exp{\left[F_2(\mathcal{E}_2)+\frac{1}{2}F_2''(\mathcal{E}_2)\mathcal{E}^2\right]}\nonumber\\
&=&  \exp{\left[F_2(\mathcal{E}_2)\right]}\left\lbrace \sqrt{\frac{2\pi}{| F_2''(\mathcal{E}_2)|}}+ \sqrt{\frac{2\pi}{F_2''\left(\mathcal{E}_2\right)}}\text{Erf}\left(\sqrt{\frac{F_2''\left(\mathcal{E}_2\right)}{2}}\frac{V}{4}\frac{s'(E/V)}{|s''(E/V)|}\delta\right)\right\rbrace \nonumber\\
&= & \exp{\left[F_2(\mathcal{E}_2)\right]}\sqrt{\frac{2\pi}{| F_2''(\mathcal{E}_2)|}}\left(1 + \frac{V}{2}\frac{s'(E/V)}{|s''(E/V)|}\sqrt{\frac{|F_2''(E/V)|}{2\pi}}\delta +\mathcal{O}(\delta^2)\right)
\eea 
where in the last step we used the approximation to the error function: $\text{Erf}(x)\approx \frac{2x}{\sqrt{\pi}}+\mathcal{O}(x^2)$. Therefore: 
\bea
S^{\text{dom}}_n-S^{\text{MC}}_n &=& \frac{1}{1-n}\ln{\left\lbrace \frac{\int d\mathcal{E}\; \exp{\left[F_{\text{dom}}(\mathcal{E})\right]}}{\int d\mathcal{E}\; \exp{\left[F_2(\mathcal{E})\right]}}\right\rbrace} \nonumber\\
&=&- \sqrt{V}\frac{s'(E/V)}{\sqrt{|s''(E/V)|2\pi}}+\mathcal{O}(\delta)
\eea
This is the $\mathcal{O}\left(\sqrt{V}\right)$ enhanced correction at the transition point $f=1/2$, now manifested directly in Renyi entropy for Renyi index $n$ satisfying (\ref{eq:enhancement_regime}). We point out that although $F_{\text{dom}}\left(\mathcal{E}\right)$ has a ``cusp'' singularity at $\mathcal{E} = E/2$, it is regularized by the correction term $F_\Delta \left(\mathcal{E}\right)$, making $F_1\left(\mathcal{E}\right)$ a smooth function across it; when the $\mathcal{O}\left(\sqrt{V}\right)$ enhanced correction appears, $F_\Delta$ fills the cusp dip between the two colliding saddles of $F_1$, creating an approximate flat interval of width $\mathcal{O}(n-1)$ between the saddles (see Figure~\ref{fig:F_plots}). In some sense, one may understand the enhanced correction as due to the effective ``soft mode'' associated with the flat interval. 

The result is expressed in terms of a general density of states. For illustration we can plug in the Cardy formula for 2 dimensional CFTs: 
\be
S(E)=2\pi\sqrt{\frac{c}{6}L_0} = 2\pi V \sqrt{\frac{c E}{6V}},\;\;s(E/V)=2\pi \sqrt{\frac{c E}{6V}}
\ee
The correction to microstate Renyi entropy takes the form: 
\be
\overline{S_n}-S^{\text{MC}}_n =  -\sqrt{V T c} +\mathcal{O}(\delta)
\ee
for $\delta \ll (c L_0)^{-1/4}$, where $T$ is the effective temperature of the global state through the relation $T V \propto \sqrt{\frac{L_0}{c}}$. 

More generally, we notice that the enhanced correction and its regime of validity are both controlled by a characteristic (dimensionless) ratio: 
\be
\gamma = \frac{Vs'(E/V)}{\sqrt{| s''(E/V) |}}.
\ee
Using standard thermodynamic relations: $\frac{1}{T}=\frac{dS}{dE},\;C_V = \frac{dE}{dT}$, we find that this ratio is indeed simply the square root of specific heat: $\gamma = \sqrt{C_V}$. We conclude that for $n-1\ll 1/\sqrt{C_V}$, the microstate Renyi entropy differs from the micro-canonical ensemble by: 
\be
\overline{S_n}-S^{\text{MC}}_n = -\sqrt{\frac{C_V}{2\pi}}+\mathcal{O}\left(\delta\right)  
\ee
This is consistent with the enhanced correction to the von Neumann entropy computed in \cite{Srednicki:2019}. 

\section{Black hole microstates in AdS/CFT}\label{sec:holography}

In the previous section, starting from the ansatz (\ref{eq:random_z}) for chaotic high energy eigenstates, we have analyzed the emergence of $\mathcal{O}\left(\sqrt{V}\right)$ enhanced correction to the Renyi entropy near the entanglement transition $V_A/V=1/2$. In this section, we study the same phenomenon in the context of AdS/CFT, where entanglement entropy can be computed geometrically by the RT surface area~\cite{Ryu2006,Ryu2006_2} and its dynamical HRT generalization~\cite{Hubeny_2007}.  The phenomenon corresponds to the enhanced correction near the phase transition in the RT surface~\cite{Headrick2010, Hartman2013, Faulkner2013} in high energy/temperature black hole microstates. 

\subsection{Boundary Euclidean path-integrals}

We begin with the assumption that a black hole microstate can be represented by a typical random state built from superposition of eigenstates in a narrow energy window around $E$: 
\be
|E,\hat{c}\rangle \propto \sum_{E-\Delta<E_i <E+\Delta} c^i |E_i\rangle
\ee

We put in an additional label $\hat{c}$ to denote a particular random choice of $c^i$ for the microstate. To proceed, let us be more explicit about the Euclidean path-integral representation for the reduced density matrix on the subsystem $A$ in the chaotic microstate:
\be
\rho_A\left(\hat{c}\right)=\text{tr}_{\hat{A}}|E,\hat{c}\rangle\langle E, \hat{c}|
\ee 

\begin{figure}[h!]
\centering
\includegraphics[scale=0.35]{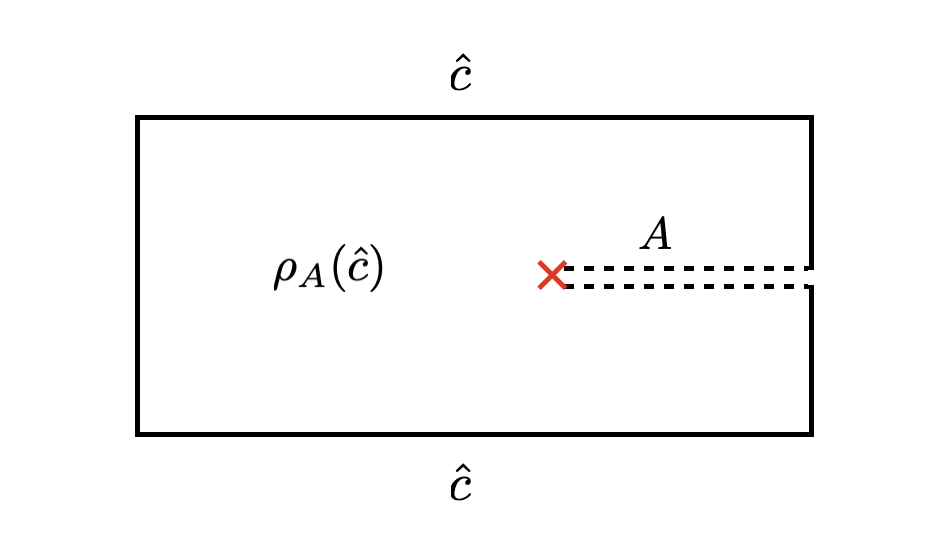}  
\includegraphics[scale=0.43]{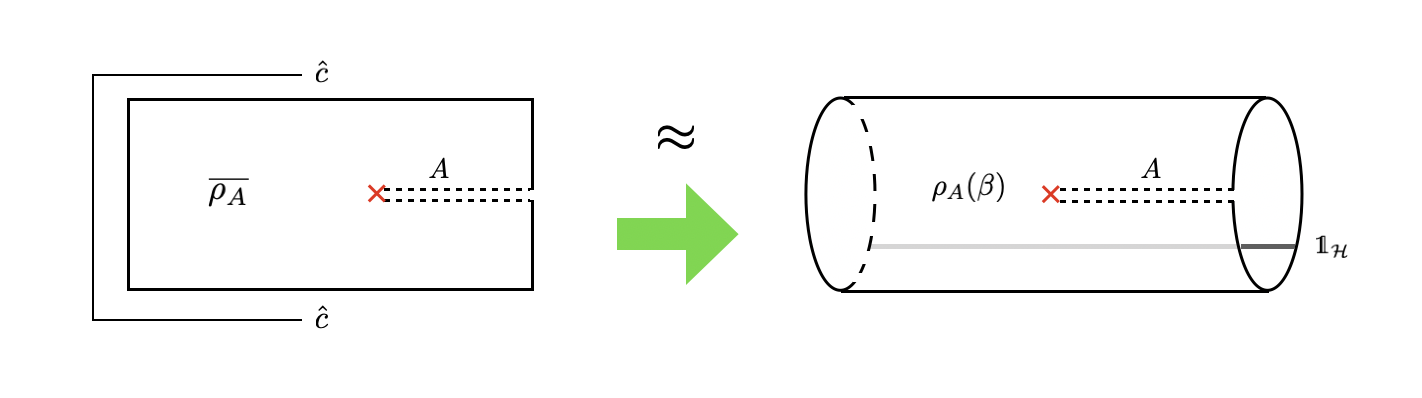}  
\caption{Left: Euclidean path-integral for the reduced density matrix $\rho_A(\hat{c})$ from a particular chaotic microstate; right: emergent KMS condition for averaged $\hat{\rho_A}$
}
\label{fig:density_matrix}
\end{figure}

This can be represented by a Euclidean path-integral on a stripe with open slit along $A$, where the random parameters $\hat{c}$ are specified at the boundary, see Figure~\ref{fig:density_matrix}. Averaging over a pair of randomness amounts to the replacement: 
\be
\overline{\hat{c}_{\text{top}}\;\hat{c}_{\text{bottom}}}\rightarrow \sum_{E-\Delta< E_i< E+\Delta} | E_i \rangle_{\text{top}} \langle E_i | _{\text{bottom}} \approx \mathds{1}_{\mathcal{H}}
\ee

This line might require a little clarification. Strictly speaking, for a random state taken from the energy window around $E$, disorder average via Wick contraction only gives rise to a projection operator into the energy window: $\overline{\hat{c}_{\text{top}}\;\hat{c}_{\text{bottom}}}\rightarrow\hat{P}_E$, i.e.  micro-canonical ensemble, instead of the full identity $\mathds{1}_{\mathcal{H}}$. The agreement between $\hat{P}_E$ and $\mathds{1}_{\mathcal{H}}$ is optimized by evolving the chaotic state in Euclidean time:
\be
|E,\hat{c}\rangle \to e^{-\frac{\beta}{2}\hat{H}}|E,\hat{c}\rangle 
\ee
and do the random averaging. Then we have:
\be 
e^{-\beta E}\hat{P}_E \approx \sum_{E'} e^{-\beta E'}|E'\rangle\langle E'| = e^{\beta \hat{H}}\mathds{1}_{\mathcal{H}}
\ee
where $\beta$ is the effective temperature fixed by the Laplace transformation from micro-canonical ensemble to canonical ensemble. In other words, if one wish to replace the disorder Wick contraction by an identity operator that glues the Euclidean path-integrals, the corresponding width needs to be dynamically fixed, see Figure~\ref{fig:density_matrix}. This is how KMS condition emerges for a chaotic high energy pure state. 

Let us now compute the trace of powers of the pure state density matrix: 
\be
\text{tr}\rho^n_A = \text{tr}_A \left\lbrace\text{tr}_{\bar{A}}|E,\hat{c}\rangle \langle E,\hat{c}|\right\rbrace^n 
\ee

\begin{figure}[h!]
\centering
\includegraphics[scale=0.29]{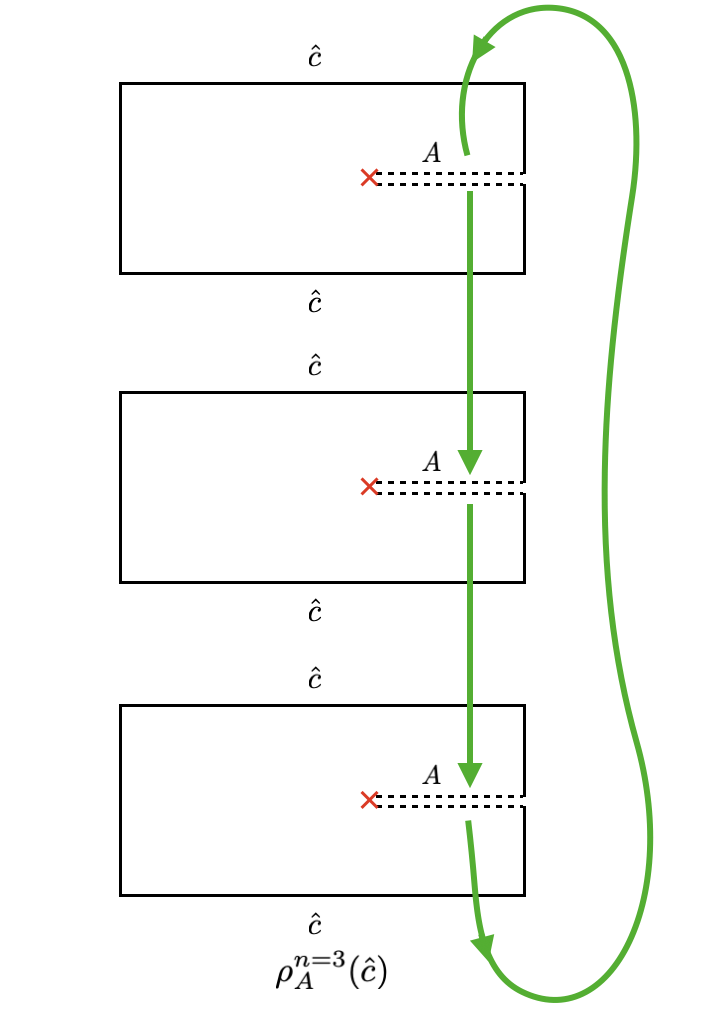}  
\includegraphics[scale=0.29]{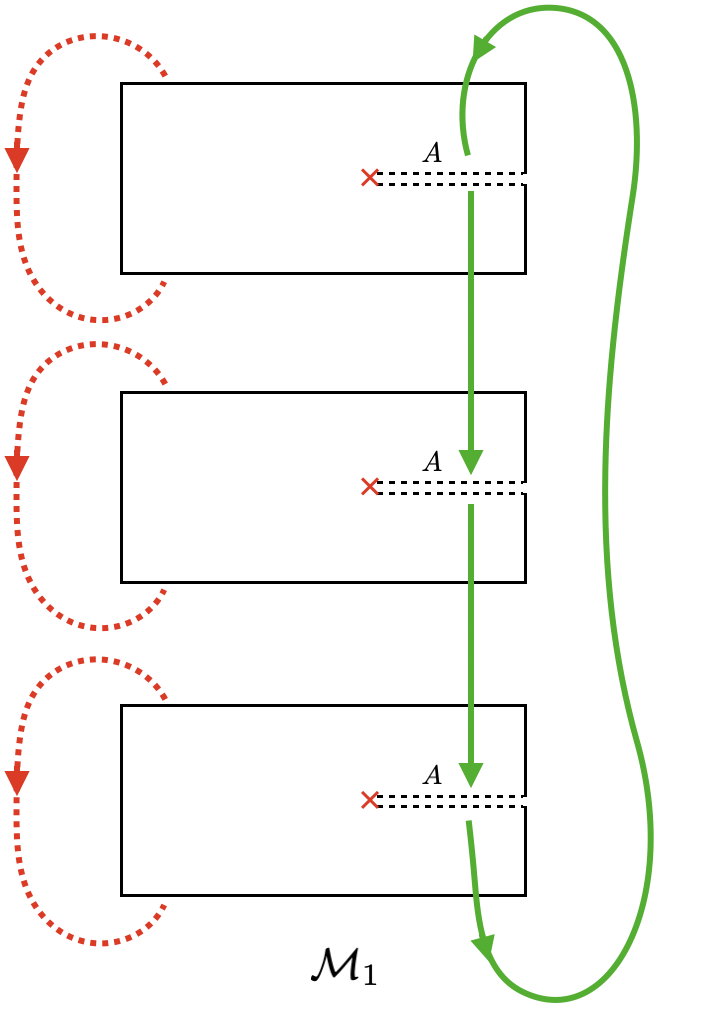}  
\includegraphics[scale=0.29]{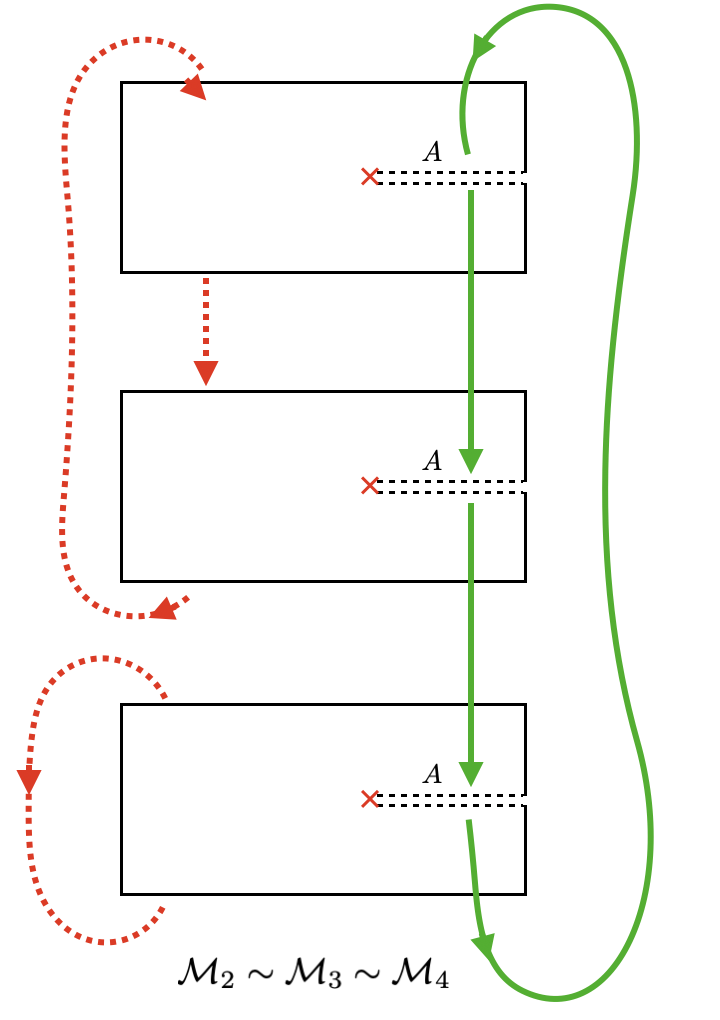}  
\includegraphics[scale=0.29]{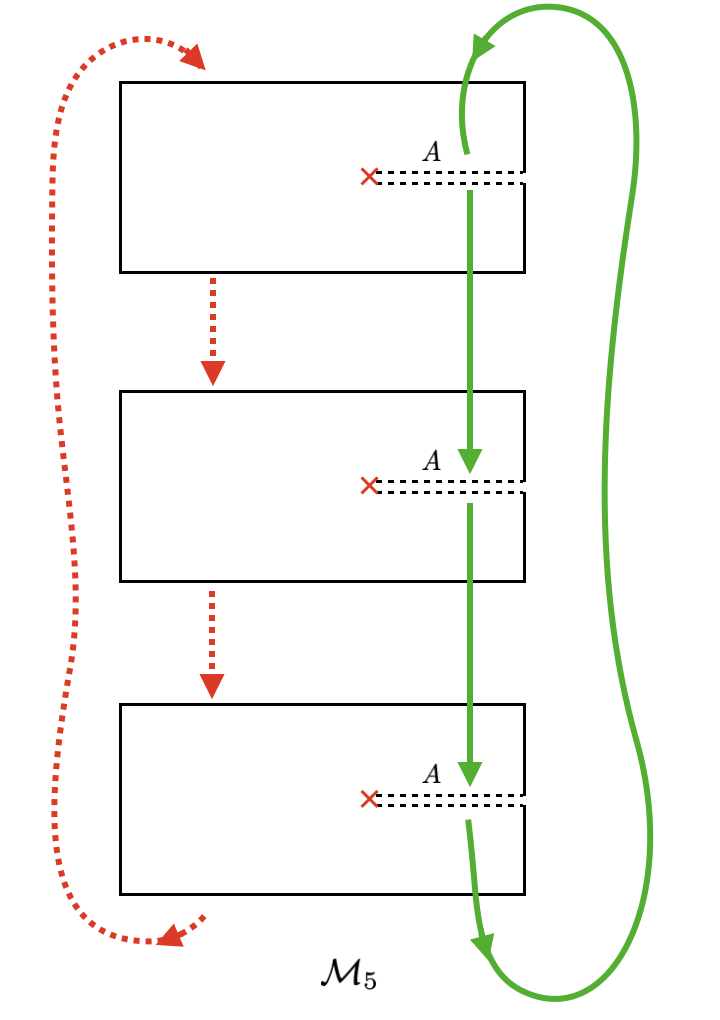}  
\caption{The four closed manifolds $\mathcal{M}_i, i=1,..,5$ that emerge after disorder averaging $\hat{c}$ in $\rho_A^n(\hat{c})$ for $n=3$.
}
\label{fig:rho3}
\end{figure}

This can be represented by the Euclidean path-integral on the branched manifold with open boundaries $\mathcal{M}_{\hat{c}_1,\hat{c}_2,...,\hat{c}_{2n}}$ (see Figure~\ref{fig:rho3}), where $\hat{c}_i, i=1,...,2n$ are $2n$ copies of the random variables specified at the boundaries of Euclidean path-integrals for the $n$ bra's $|E,\hat{c}\rangle$ and the $n$ ket's $\langle E,\hat{c}|$. Upon disorder averaging, these random variables will pair-up into $n$ approximately identity operators (after appropriate evolution in Euclidean time) that sew the corresponding boundaries. After the gluing, different ways of pairing up the random variables give rise to path-integrals on different branched manifolds that are all closed. Schematically: 
\be 
\text{tr}\rho^n_A = \frac{Z_n}{Z_1^n},\quad
Z_n = \int \mathcal{D}\phi\;e^{-I_E\left(\mathcal{M}_{c_1,...,c_{2n}},\phi\right)}\approx \sum_{\mathcal{M}_i} \;\int \mathcal{D}\phi\;e^{-I_E\left(\mathcal{M}_i,\phi\right)}
\ee

They correspond to the different terms when performing Wick contraction in Sec.~\ref{sec:general} with the chaotic ansatz,  see Eq.~\ref{eq:wick_contraction}. Similar to there, the dominant contributions come from the ``planar'' contractions. 

\begin{figure}[h!]
\centering
\includegraphics[scale=0.45]{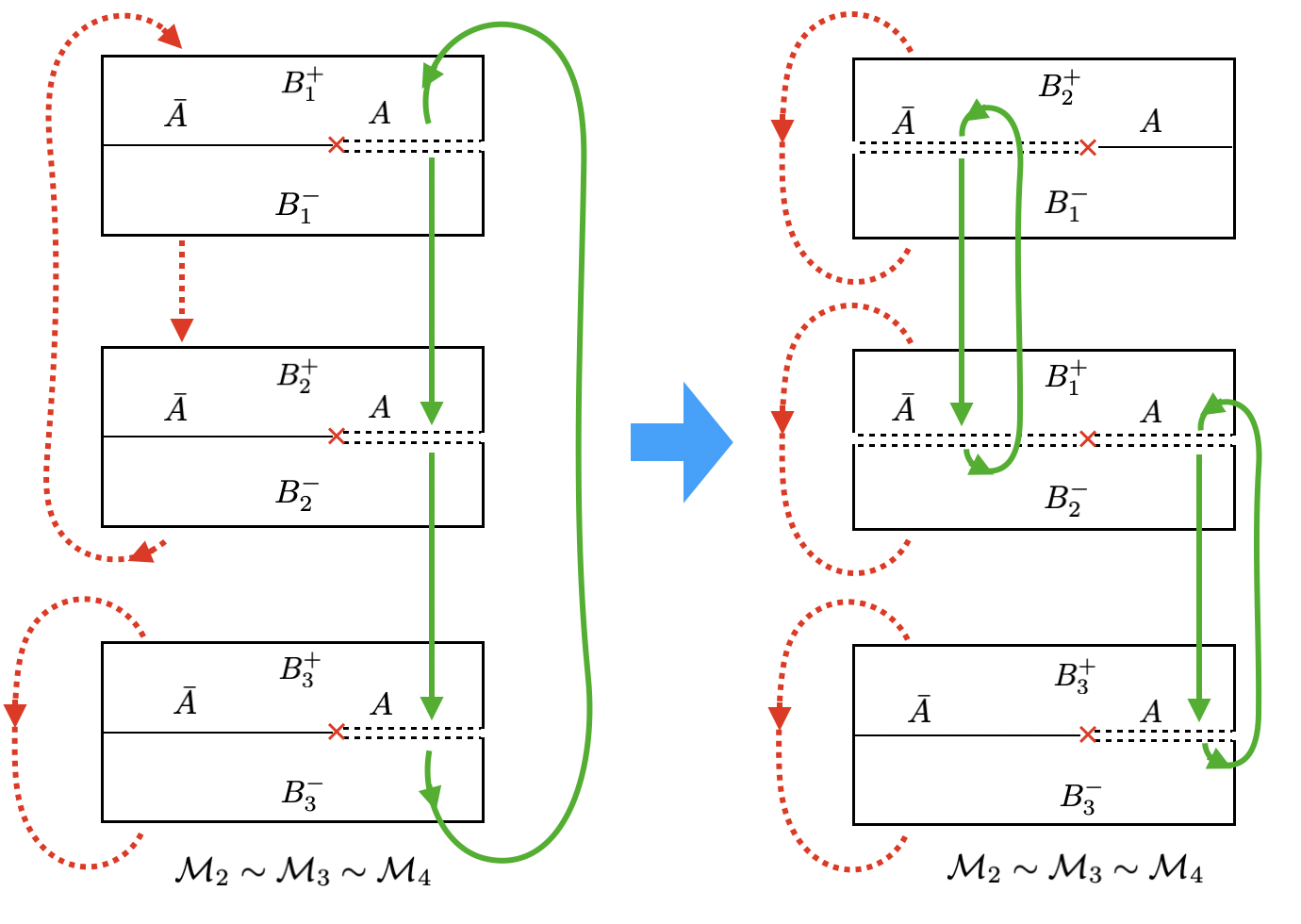}  
\includegraphics[scale=0.45]{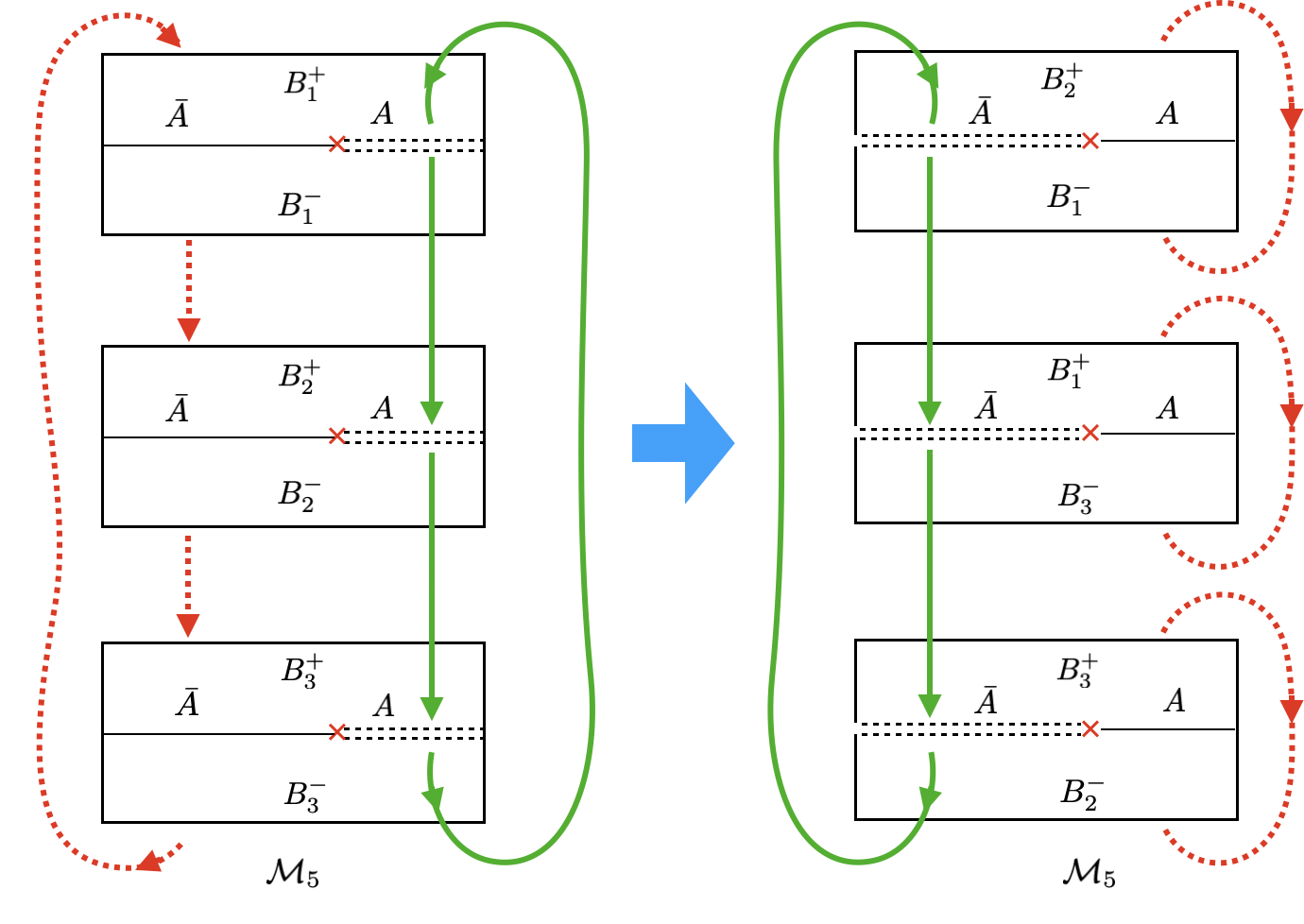}  
\caption{``Canonical'' representations of $\mathcal{M}_2\sim \mathcal{M}_3\sim \mathcal{M}_4$ and $\mathcal{M}_5$ for $n=3$.
}
\label{fig:cano}
\end{figure}

For each $\mathcal{M}_i$ obtained from ``planar'' contractions, if we re-arrange the orders of the gluing procedures, we can translate all $\mathcal{M}_i$'s into ``canonical'' representations (see Figure~\ref{fig:cano}), where the Wick contractions become the same: they only pair up adjacent ``bra'' and ``ket'' boundaries, i.e: 
\be 
|E,\hat{c}\rangle \langle E,\hat{c}| \to e^{-\beta \hat{H}}\mathds{1}_{\mathcal{H}} \propto \rho_{\mathcal{H}}(\beta)
\ee

To compensate for this uniformity in contraction patterns, the branching structure is different for different $\mathcal{M}_i$'s. For example, the contraction $(1\leftrightarrow 2)...(2n-1 \leftrightarrow 2n)$ coincides with the canonical one, so the branching structure is the same: $\left(\bar{A}_1\leftrightarrow \bar{A}_2\right)...\left(\bar{A}_{2n-1}\leftrightarrow \bar{A}_{2n}\right)$ on $\bar{A}$ and $\left(A_2\leftrightarrow A_3\right)...\left(A_{2n}\leftrightarrow A_{1}\right)$ on A; on the other hand, the contraction $(2\leftrightarrow 3)...(2n \leftrightarrow 1)$ gives rise to the following branching structure after changing into the canonical representation: $\left(\bar{A}_2\leftrightarrow \bar{A}_3\right)...\left(\bar{A}_{2n}\leftrightarrow \bar{A}_1\right)$ on $\bar{A}$ and $\left(A_1\leftrightarrow A_2\right)...\left(A_{2n-1}\leftrightarrow A_{2n}\right)$ on A, see $\mathcal{M}_5$ in Figure~\ref{fig:cano}.  The notations should be self-explanatory. The later contraction gives rise to a path-integral that computes the thermal Renyi entropy for $\bar{A}$, this term is responsible for the second part of the pure state ``Page curve'', and is analogous to the replica wormholes studied in \cite{replica_wormhole_east,replica_wormhole_west}. It is easy to see that for contractions other than the two special cases, the canonical representation will contain ``mixed'' branch structures, see $\mathcal{M}_{2,3,4}$ in Figure~\ref{fig:cano}. 

\subsection{Bulk Euclidean saddle points}

Now let us extend into the bulk in AdS/CFT. Before disorder average, the boundary state $|E,\hat{c}\rangle$ is expected to be dual to a high energy black hole microstate with an end-of-world (EoW) brane inserted behind the horizon. The boundary conditions on the EoW brane is specified by the random variables $\hat{c}$. After disorder average, the gluing on the boundary is extended into bulk, i.e. the EoW branes are also glued pairwise. As a result, each of the boundary Euclidean path-integrals on $\mathcal{M}_i$ is dominated by a corresponding bulk saddle point $\mathcal{B}_i$ with $\partial\mathcal{B}_i = \mathcal{M}_i$, there is no more EoW branes.

 In general, finding the bulk solutions filling the interior of $\mathcal{M}_i$'s subject to boundary conditions is difficult. Fortunately, we are considering the high energy density limit, as a result the bulk Euclidean saddles are characterized by a small radial extension compared to the system size $\lbrace 0\leq r\leq \beta\rbrace,\beta \ll V_A, V_{\bar{A}}$. We allow the radius to be position dependent in general. Furthermore, we focus only on features that scale with (powers of) the subsystem volume $V_A$. This allows us to construct bulk solutions simply by gluing portions of (possibly distinct) black hole geometries whose asymptotic boundaries coincide with $A$ and $\bar{A}$ respectively, this is similar to the approach taken in \cite{Dong:2019}. The matching details at the junctions are not important for our purposes. 

\begin{figure}[h!]
\centering
\includegraphics[scale=0.5]{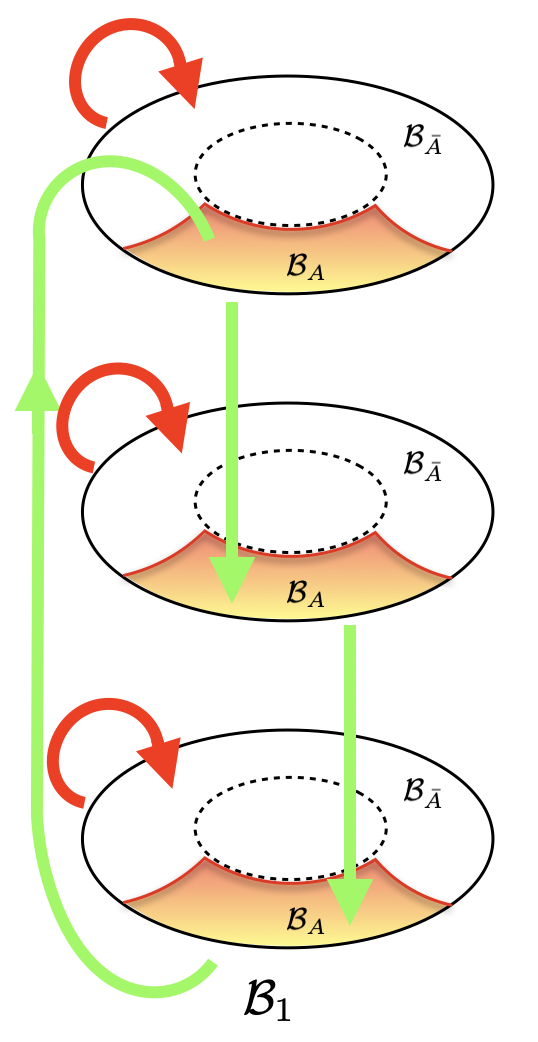}  
\includegraphics[scale=0.5]{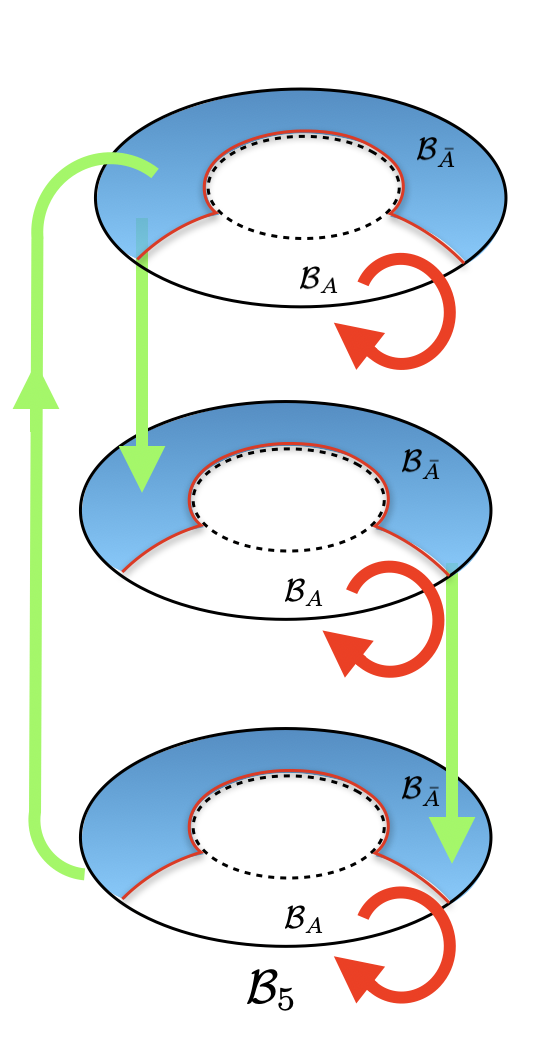}  
\includegraphics[scale=0.5]{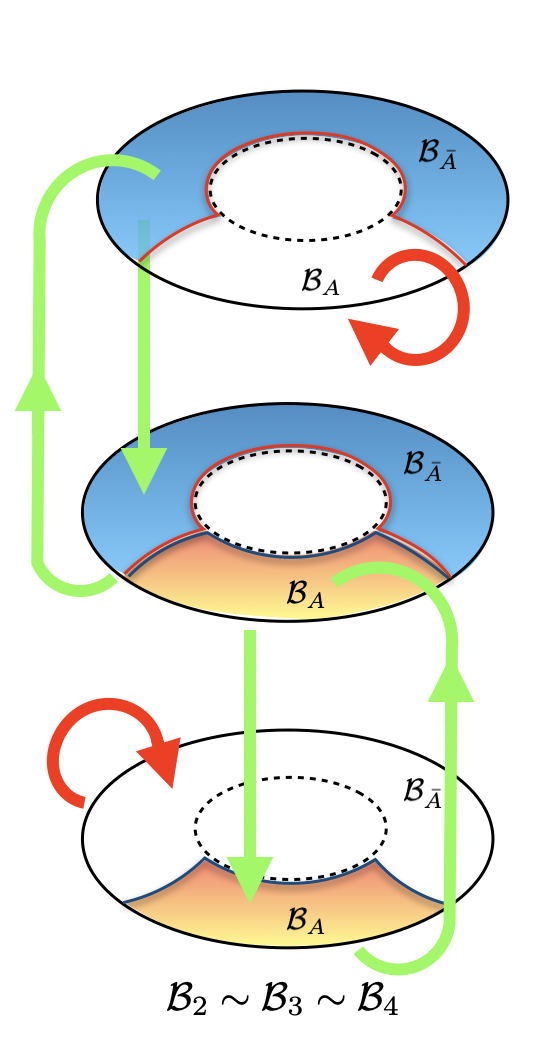}  
\caption{``Kinematics'' of the bulk saddles $\mathcal{B}_i$'s corresponding to $\mathcal{M}_i$'s. Green and red arrows indicate how the three bulk time-slices are glued together for each configuration. The dotted circles represent the horizons.
}
\label{fig:bulk}
\end{figure}

For the contractions $(2\leftrightarrow 3)...(2n \leftrightarrow 1)$ and $(1\leftrightarrow 2)...(2n-1 \leftrightarrow 2n)$, there is only a unique black hole geometry along each copy of $A$ and $\bar{A}$ respectively, and replica-symmetry is preserved in these saddles; for other choices of Wick contractions, replica-symmetry is broken and there may be multiple choices of black hole geometries along different copies of $A$ and $\bar{A}$; see examples of $n=3$ in Figure~\ref{fig:bulk}.

Having specified the ``kinematic'' part of the bulk solutions, we need to understand the dynamics, i.e. satisfying the bulk equations of motion. In the ansatz we are working with, i.e. gluing black hole geometries, the equations of motion take the form of matching conditions. There are two related perspectives for such matching conditions. We can label the black hole geometries by the corresponding inverse temperature $\beta$, then the matching condition is then simply the micro-canonical boundary condition: total energy of the microstate is $E$. For example, the matching equation of motion for the saddles corresponding to $\mathcal{M}_i, i=1,...,5$ for computing $n=3$ are:\footnote{One might find out that the solution for $\mathcal{M}_{2,3,4}$ corresponds to $\beta_1=\beta_3=\beta(E)$, $\beta_2=0$. This does not give a singular manifold, since the corresponding thermal circle for $\beta_2$ does not close; it only serves to connect.}
\bea 
&&\mathcal{M}_1:\;\;\;\;\; f E_{3\beta} + (1-f)E_\beta=E \nonumber\\
&&\mathcal{M}_{2,3,4}:\, f E_{\beta_1+\beta_2} + (1-f)E_{\beta_1}=f E_{\beta_1+\beta_2} + (1-f)E_{\beta_2+\beta_3}=f E_{\beta_3} +(1-f)E_{\beta_2+\beta_3}=E\nonumber\\
&&\mathcal{M}_5:\;\;\;\;\, f E_{\beta} + (1-f)E_{3\beta}=E 
\eea
where $E_\beta$ is the total ADM energy of the black hole at inverse temperature $\beta$. Correspondingly we can also label the black hole solutions by the total energy $E'$, now the matching condition corresponds to imposing that the total Euclidean width on $A$ and $\bar{A}$ must be equal: 
\be \label{eq:matching_beta}
\sum \beta_A(E')=\sum \beta_{\bar{A}} \left(\frac{E-fE'}{1-f}\right)
\ee
where $\beta(E')$ is the inverse temperature of black hole having total ADM energy $E'$. For example when $n=3$ we have: 
\bea
&&\mathcal{M}_1:\;\; \beta(E')=3\beta\left(\frac{E-fE'}{1-f}\right) \nonumber\\
&&\mathcal{M}_{2,3,4}:\;\; 2\beta(E')=2\beta\left(\frac{E-fE'}{1-f}\right) \nonumber\\
&&\mathcal{M}_5:\;\; 3\beta(E')=\beta\left(\frac{E-fE'}{1-f}\right)
\eea
Imposing these matching conditions ensures the absence of conical defects in the \textit{full} bulk geometries $\mathcal{M}_i,i=1,2,3$. This is the only remaining content of Einstein's equations in the regime and ansatz we are working with. Indeed they would correspond to the saddle point equations we could impose on each term individually in the expansion Eq.~\ref{eq:trace_intergal}.

\subsection{Re-summing saddles: cosmic brane effective action}
As has been observed in Sec.~\ref{sec:general}, in order to study the details near the transition point we should consider all the replica-nonsymmetric configurations, re-sum them into an effective action which we then compute semi-classically as a whole. By now we have understood what these configurations correspond to holographically, let us proceed with the re-summation in the context of AdS/CFT.

We leave the subsystem energies unfixed, but still impose the boundary condition for the total energy $E_A + E_{\bar{A}}=E$. Holographically this means that we do not fix the black hole geometries involved for the $\mathcal{M}_i$'s. Denote the portion of the black hole geometry extending along $A$ by $\mathcal{B}_A(E')$, and the corresponding semi-classical contribution to the Euclidean action by\footnote{Defining the holographic action for a bulk subsystem is very subtle; however for the regime and limit of our interest here, such subtlety is subdominant.}
\be
I_A(E')=\int_{\mathcal{B}_A(E')} \left(\mathcal{L}_{\text{E.H.}}+\mathcal{L}_{\text{matter}}\right)+\int_A \mathcal{L}_{\text{H.G.}} 
\ee
where the bulk Lagrangian densities $\left\lbrace \mathcal{L}_{\text{E.H.}},\mathcal{L}_{\text{matter}},\mathcal{L}_{\text{H.G}}\right\rbrace$ are evaluated at the saddle point geometry $\mathcal{B}_A(E')$, as well as matter field configurations not specified here. Consequently, let us also denote the portion of black hole geometry and semi-classical Euclidean action along $\bar{A}$ by $\mathcal{B}_{\bar{A}}\left(\frac{E-fE'}{1-f}\right)$ and $ 
I_{\bar{A}}\left(\frac{E-fE'}{1-f}\right)$ respectively, this is obtained after imposing the total energy conditions.  

\begin{figure}[h!]
\centering
\includegraphics[scale=0.31]{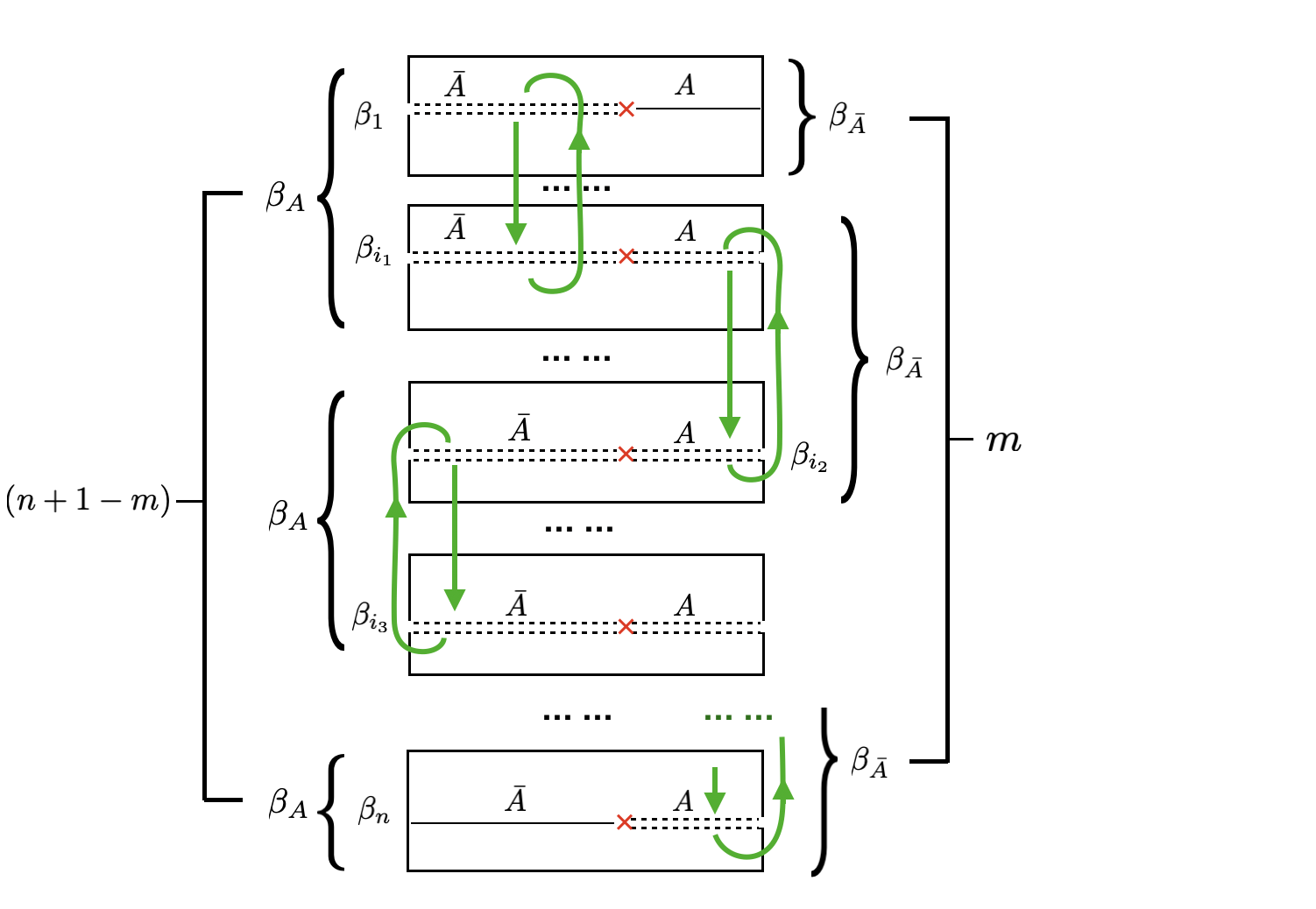}  
\includegraphics[scale=0.28]{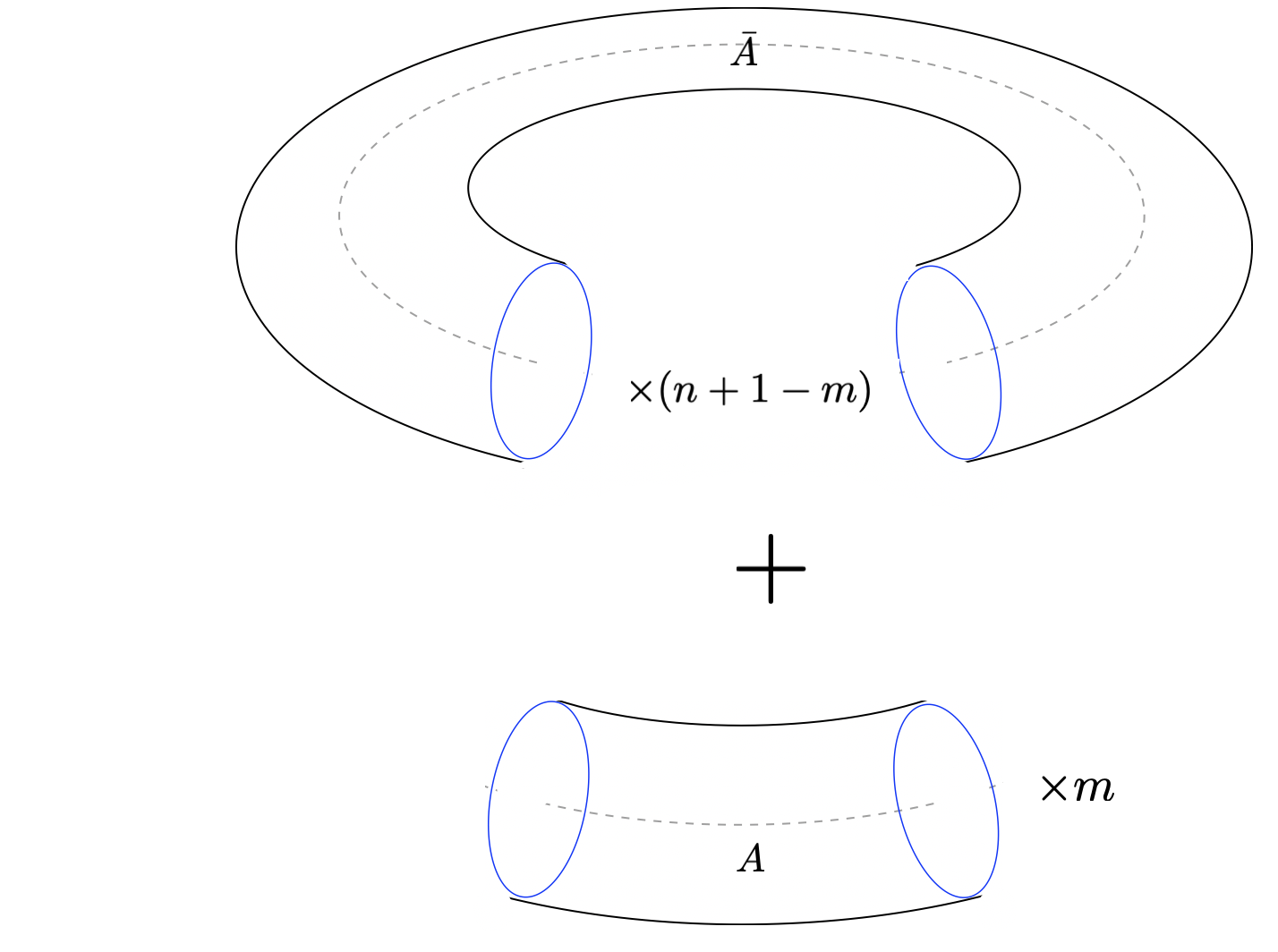}  
\includegraphics[scale=0.31]{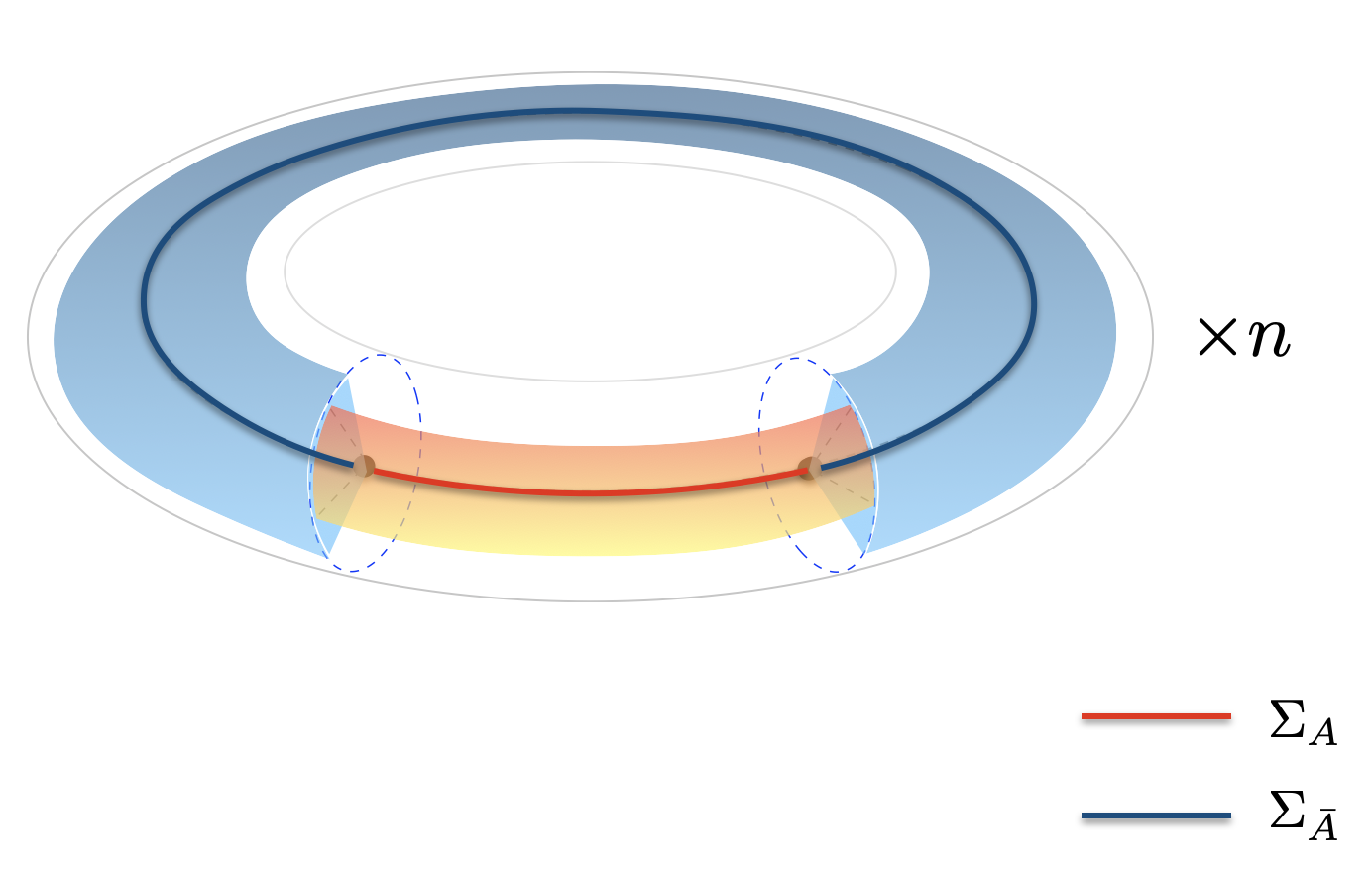}  
\caption{Top-left: a generic planar $\mathcal{M}_i\in\left\lbrace \mathcal{M}_m\right\rbrace$. Top-right: a generic bulk saddle $\mathcal{B}_m$ for $\left\lbrace \mathcal{M}_m\right\rbrace$. Bottom: the quotient geometry $\tilde{\mathcal{B}}_m = \mathcal{B}_m/\mathbb{Z}_n$, obtained by the double-defect construction. 
}
\label{fig:bulk_quot}
\end{figure}

Using these ingredients, the total semi-classical Euclidean action for each of the $\mathcal{M}_i$'s takes the form: 
\be\label{eq:bulk_action_m}
I^n_{\mathcal{M}_i}(E) = \text{Min} \left\lbrace  m I_A(E') + (n+1-m)I_{\bar{A}}\left(\frac{E-fE'}{1-f}\right): E'\right\rbrace
\ee
This corresponds to $\mathcal{M}_i$ obtained from gluing $m$ copies of black hole portion $\mathcal{B}_A(E')$ with $(n+1-m)$ copies of $\mathcal{B}_{\bar{A}}\left(\frac{E-fE'}{1-f}\right)$; see the top of Figure~\ref{fig:bulk_quot}. We have not specified the boundaries of $\mathcal{B}_A(E')$ or $\mathcal{B}_{\bar{A}}\left(\frac{E-fE'}{1-f}\right)$ in the bulk, and in general directly gluing them would result in discontinuous junctions across these boundaries. As commented before, these issues do not enter in the limit we are interested. We label the boundary and bulk manifolds by $\lbrace \mathcal{M}_m\rbrace $ and $\lbrace \mathcal{B}_m \rbrace$ respectively. This is not a one-to-one correspondence between the label $m$ and the original label for the contractions $i$, because there are in general multiple ways to junction $m$ and $(n+1-m)$ portions of black hole geometries by gluing the original EoW branes. For this reason we include a curly bracket to indicate that each $\lbrace \mathcal{M}_m\rbrace$ and $\lbrace \mathcal{B}_m\rbrace$ represent the class of all contraction or gluing choices resulting in the same $m$. 

For each class $\lbrace\mathcal{M}_m\rbrace$, the bulk saddles in $\lbrace \mathcal{B}_m\rbrace$ can be constructed in a way analogous to the cosmic brane prescription in \cite{Dong:2016}. We describe the construction as follows. For each $\lbrace \mathcal{B}_m\rbrace$, the saddle consists of $n$ identical wedges $\tilde{\mathcal{B}}_m$,
\be 
\tilde{\mathcal{B}}_m = \mathcal{B}_m/\mathbb{Z}_n
\ee
Each wedge $\tilde{\mathcal{B}}_m$ is then constructed by inserting in the original black hole state $|E\rangle_{\text{bulk}}$ a pair of defects consisting of two cosmic branes $\Sigma_A$ and $\Sigma_{\bar{A}}$, homologous to $A$ and $\bar{A}$ respectively. For a particular $m$ their brane tensions $T_A$ and $T_{\bar{A}}$ take the values:  
\be
T_A = \frac{n-m}{4nG_N},\;\;T_{\bar{A}}= \frac{m-1}{4nG_N}
\ee
respectively. Backreaction from these defects will result in conical singularities with angular extensions: 
\be
\Delta \phi_A = 2\pi\frac{m}{n},\;\;\Delta \phi_{\bar{A}} = 2\pi \frac{n+1-m}{n} 
\ee
for the portions of $\tilde{\mathcal{B}}_m$ along $A$ and $\bar{A}$ respectively. The union $\mathcal{B}_m$ of $n$ such wedges will therefore be smooth and satisfy the junction condition that computes (\ref{eq:bulk_action_m}). We name it the \textit{double-defect} construction (see Figure~\ref{fig:bulk_quot}).

Let us make a few comments regarding the double-defect construction. It may appear that we have proposed a procedure based on the replica-symmetry among the $n$ wedges $\tilde{\mathcal{B}}_m$. However, we can do this only because there is an effective $U(1)$ rotation-symmetry along the bulk thermal circle that emerges in the limit of our interest: high energy $E\ell \gg 1$ and ignoring boundary effects near $\partial{A}$. It allows the division of the circle into $n$ equal wedges that are therefore replica-symmetric under $\mathbb{Z}_n \subset U(1)$, justifying our construction. The origin for this emergent rotation-symmetry can be traced to two ingredients. Firstly, the $m$ copies of black hole portions along $A$ are all the same (and similarly for $\bar{A}$), we assume this to be the dynamically favored configuration. More importantly, we are approximating each of the $m$ black hole portions as ``featureless'' and thus invariant under rotation along the bulk thermal circle. This is certainly not true in the exact solution. For example, solving the matching equations will in general give a set $\lbrace \beta_1, ..., \beta_n\rbrace$ that are not all equal; see top-left of Figure~\ref{fig:bulk_quot}. Once we include the details arising from gluing and smoothening across $\partial{A}$, they will produce ``features'' along the thermal circles that generically breaks the rotation-symmetry. Apart from this, these features also distinguish between the bulk saddles $\mathcal{B}_i$'s in the same class $\lbrace \mathcal{B}_m\rbrace$, because different contractions $i$ will in general give different sets for $\lbrace \beta_1, ..,\beta_n\rbrace$, and thus different gluing and smoothening effects. Since these features are localized near $\partial A$, in the high energy limit they only affect the radial-going portion of the bulk defects that we have ignored. It is interesting to consider the corrections they induce to the cosmic brane picture, we leave this for future investigations. 

One way to represent the double-defect construction is to simply add the corresponding brane source terms into the bulk action. The total on-shell action $I_{\text{bulk}}\left(\mathcal{B}_m\right)$ can be divided into $n$ quotient on-shell actions:
\be\label{iquotient}
I_{\text{bulk}}\left(\mathcal{B}_m\right) = n I_{\text{bulk}}(\tilde{\mathcal{B}}_m)
\ee
with the quotient on-shell action $I_{\text{bulk}}(\tilde{\mathcal{B}}_m)$ given by a quotient path integral in the saddle point approximation:
\be\label{eq:quotient-saddle}
e^{-I_{\text{bulk}}(\tilde{\mathcal{B}}_m)} =  \int \mathcal{D}g\mathcal{D}\phi \mathcal{D}\Sigma_A \mathcal{D}\Sigma_{\bar{A}}\;\;  e^{-I_{\text{bulk}}\left(g,\phi,E\right)-\frac{n-m}{4nG_N} I_{\text{brane}}\left(\Sigma_A\right)-\frac{m-1}{4nG_N}I_{\text{brane}}\left(\Sigma_{\bar{A}}\right)}
\ee 
where $I_{\text{bulk}}\left(g,\phi,E\right)$ is the bulk Euclidean action with boundary conditions specified by the original state $|E\rangle$, and $I_{\text{brane}}\left(\Sigma_A\right)$, $I_{\text{brane}}\left(\Sigma_{\bar{A}}\right)$ are the Nambu-Goto actions for cosmic branes located on $\Sigma_A$, $\Sigma_{\bar{A}}$ respectively:
\be
I_{\text{brane}}\left(\Sigma_A\right) = \int_{\Sigma_A} d^{d-1} y \sqrt{\gamma},\;\;I_{\text{brane}}\left(\Sigma_{\bar{A}}\right) =  \int_{\Sigma_{\bar{A}}} d^{d-1} y \sqrt{\gamma}.
\ee
Here $\gamma$ is the induced metric on the brane.  Although not shown explicitly, $I_{\text{brane}}$ depends implicitly on the spacetime metric $g$.  In Eq.~\eqref{eq:quotient-saddle}, $\mathcal{D}\Sigma_A$ denotes a path integral over the location of the brane.

Extremizing over $\lbrace g,\phi, \Sigma_A, \Sigma_{\bar{A}}\rbrace$ solves for the backreacted geometry for a particular saddle $\mathcal{B}_m$. As was alluded to, our strategy is to re-sum all the saddle configurations before extremizing. Before proceeding, we point out that while (\ref{eq:quotient-saddle}) computes the action of the $\mathbb{Z}_n$ quotient $\tilde{\mathcal{B}}_m$ of the ``parent space'' $\mathcal{B}_m$, it is the full action of the parent space $\mathcal{B}_m$ that we need to re-sum. In the semi-classical limit, we can simply write the latter as
\be 
e^{-I_{\text{bulk}}\left(\mathcal{B}_m\right)} =  \int \mathcal{D}g\mathcal{D}\phi \mathcal{D}\Sigma_A \mathcal{D}\Sigma_{\bar{A}}\;\;  e^{-n I_{\text{bulk}}\left(g,\phi,E\right)-\frac{n-m}{4G_N} I_{\text{brane}}\left(\Sigma_A\right)-\frac{m-1}{4G_N}I_{\text{brane}}\left(\Sigma_{\bar{A}}\right)}
\ee
using Eqs.~\eqref{iquotient} and \eqref{eq:quotient-saddle}.

Comparing with the combinatorics factors worked out in Appendix~\ref{app:contraction}, we find that out of all the contractions $i$'s, there are $N(n,m)=\frac{1}{n}\begin{pmatrix}
n\\
m
\end{pmatrix}\begin{pmatrix}
n\\
m-1
\end{pmatrix}$ number of ``micro-configurations'' for the class $\lbrace \mathcal{M}_m\rbrace $. So we can re-sum all the contractions and obtain the full replicated partition function $Z_n$: 
\bea
Z_n &=& \sum^{n}_{m=1} N(n,m) \int \mathcal{D}g\mathcal{D}\phi \mathcal{D}\Sigma_A \mathcal{D}\Sigma_{\bar{A}}\;\;  e^{-n I_{\text{bulk}}\left(g,\phi,E\right)-\frac{n-m}{4G_N} I_{\text{brane}}\left(\Sigma_A\right)-\frac{m-1}{4G_N}I_{\text{brane}}\left(\Sigma_{\bar{A}}\right)} \nonumber\\
&=& \int \mathcal{D}g\mathcal{D}\phi \mathcal{D}\Sigma_A \mathcal{D}\Sigma_{\bar{A}}\;\;  e^{-n I_{\text{bulk}}\left(g,\phi,E\right)} \sum^{n}_{m=1} N(n,m)\;e^{-\frac{n-m}{4 G_N} I_{\text{brane}}\left(\Sigma_A\right)-\frac{m-1}{4 G_N}I_{\text{brane}}\left(\Sigma_{\bar{A}}\right)}\nonumber\\
&=& \int \mathcal{D}g\mathcal{D}\phi \mathcal{D}\Sigma_A \mathcal{D}\Sigma_{\bar{A}}\;\;  e^{-n I_{\text{bulk}}\left(g,\phi,E\right)} G_n\left(\Sigma_A,\Sigma_{\bar{A}}\right)\nonumber\\
&=& \int \mathcal{D}g\mathcal{D}\phi \mathcal{D}\Sigma_A \mathcal{D}\Sigma_{\bar{A}}\;\;  e^{-n I_{\text{bulk}}\left(g,\phi,E\right) - I_{\text{eff}}\left(\Sigma_A,\Sigma_{\bar{A}},n\right)}\label{zneff}
\eea
where we have defined
\be
G_n\left(\Sigma_A, \Sigma_{\bar{A}}\right) = \mbox{$_2$F$_1$}\left[1-n,-n;\,2;\, e^{-\frac{|\Delta I_{\text{brane}}|}{4 G_N}}\right] \times \begin{cases}
e^{-\frac{n-1}{4 G_N}I_{\text{brane}}(\Sigma_A)},\; \Delta I_{\text{brane}}<0\\
e^{-\frac{n-1}{4 G_N}I_{\text{brane}}\left(\Sigma_{\bar{A}}\right)},\; \Delta I_{\text{brane}}>0
\end{cases}
\ee
with
\be
\Delta I_{\text{brane}} \equiv I_{\text{brane}}\left(\Sigma_A\right)-I_{\text{brane}}\left(\Sigma_{\bar{A}}\right),
\ee
and we have extracted a total ``effective action'' for the cosmic branes
\bea \label{eq:Sb_eff}
&I_{\text{eff}}& \left(\Sigma_A, \Sigma_{\bar{A}},n\right) = -\ln{G_n\left(\Sigma_A,\Sigma_{\bar{A}}\right)}\nonumber\\
&=& \begin{cases} 
\frac{n-1}{4 G_N}I_{\text{brane}}(\Sigma_A) -\ln{\left(\mbox{$_2$F$_1$}\left[1-n,-n;\,2;\, e^{\frac{\Delta I_{\text{brane}}}{4 G_N}}\right]\right)},\; \Delta I_{\text{brane}}<0 \\
\frac{n-1}{4 G_N}I_{\text{brane}}(\Sigma_{\bar{A}}) -\ln{\left(\mbox{$_2$F$_1$}\left[1-n,-n;\,2;\, e^{-\frac{\Delta I_{\text{brane}}}{4 G_N}}\right]\right)},\; \Delta I_{\text{brane}}>0
\end{cases}
\eea
The first term in the effective action corresponds to that of a single cosmic brane in the dominant configuration; the second term re-sums the (semi-classical) corrections from the subdominant configurations. They correspond to the decomposition into $F_{\text{dom}}$ and $F_\Delta$ in Sec.~\ref{sec:general}. The resulting $I_{\text{eff}}$ has a modified dynamics in terms of brane dynamics and backreactions on the bulk geometry. The bulk physics near the transition point is encoded in such modifications, which we turn to study next. 

\subsection{Computation in fixed-area basis}
Let us now solve the dynamics of the effective action for the cosmic branes Eq.~(\ref{eq:Sb_eff}), which is our main result out of re-summing the saddles in the AdS/CFT calculations. After re-summation the effective-action becomes very non-local, i.e. it does not represent local insertion of defect sources. Computationally, it is therefore easiest to work in the basis of fixed-area states \cite{Dong:flat,Akers:2018fow,Dong:2019piw}. We quickly review the main ingredients below. In general, assuming time-reflection symmetry we can decompose a holographic bulk state schematically as follows: 
\bea\label{eq:decom_gen}
| \psi \rangle_{\text{bulk}} = \sum_{\alpha,i,j}C_{\alpha, ij}|\alpha, i\rangle_a |\alpha,j\rangle_{\bar{a}}
\eea
where $\lbrace a,\bar{a}\rbrace $ with $\partial a = \Sigma_{\text{RT}} \cup A,\; \partial{\bar{a}}=\Sigma_{\text{RT}}\cup \bar{A}$ is a partition of the bulk Cauchy slice across the RT surface. The label $\alpha$ denotes data specified on $\Sigma_{\text{RT}}$, including the induced metric on it, and a conformal structure in the normal plane. They correspond to the ``central'' degrees of freedom when decomposing the  bulk Hilbert space $\mathcal{H}_{code}$ in terms of the partition $\mathcal{H}_a$ and $\mathcal{H}_{\bar{a}}$: 
\be
\mathcal{H}_{code} = \oplus_{\alpha} \mathcal{H}_a(\alpha)\otimes \mathcal{H}_{\bar{a}}(\alpha)
\ee
In particular, one of the operators in the center algebra is the ``area operator'' $\hat{\mathcal{A}}$, whose eigenvalue is part of what $\alpha$ specifies. From this we can repackage the decomposition Eq.~(\ref{eq:decom_gen}) into fixed-area states as: 
\bea\label{eq:decom_fix}
|\psi\rangle_{\text{bulk}} = \sum_{\mathcal{A}}\left( \sum_{\alpha,i,j: \hat{A}[\alpha]=\mathcal{A}}C_{\alpha, ij}|\alpha, i\rangle_a |\alpha,j\rangle_{\bar{a}} \right)= \sum_{\mathcal{A}}\;|\tilde{\psi}_{\mathcal{A}}\rangle
\eea
These states as defined are orthogonal but not normalized, we can then define the normalized states:
\be
|\tilde{\psi}_{\mathcal{A}}\rangle = \sqrt{P(\mathcal{A})}\;|\psi_{\mathcal{A}}\rangle,\; \langle \psi_{\mathcal{A}} |\psi_{\mathcal{A}'}\rangle = \delta_{\mathcal{A},\mathcal{A}'}
\ee
where the normalization factor can be computed by the original bulk Euclidean path-integral projected into geometries $g$ having fixed extremal surface area $\hat{\mathcal{A}}[g] = \mathcal{A}$, and the original state is a weighted sum of these normalized fixed-area states: 
\be\label{eq:decom_area}
|\psi\rangle_{\text{bulk}} = \sum_{\mathcal{A}} \sqrt{P(\mathcal{A})}\;|\psi_{\mathcal{A}}\rangle,\;\;P(\mathcal{A}) =\mathcal{N}^{-1}\int \mathcal{D}g\mathcal{D}\phi \Big|_{\hat{\mathcal{A}}[g]=\mathcal{A}}\;e^{-I_{\text{bulk}}(g,\phi,E)} 
\ee
where $\mathcal{N}$ is the normalization constant for $|\psi\rangle_{\text{bulk}}$ itself:
\be
\mathcal{N}=\int \mathcal{D}g\mathcal{D}\phi\;e^{-I_{\text{bulk}}(g,\phi,E)}.
\ee

There are cases where there are more than one extremal surfaces sharing the same boundaries, e.g. the state is mixed or $A$ is not connected. For simplicity, say there are two such extremal surfaces $\left(\Sigma_1, \Sigma_2\right)$ such that: 
\be
\partial \Sigma_1 = \partial \Sigma_2 = \partial A 
\ee
For regular geometries the two extremal surfaces do not intersect except at the boundaries. As a result, the bulk Cauchy surface can be partitioned into three parts $(a,b,c)$ such that: 
\be
a\cap b = \Sigma_1,\; b\cap c = \Sigma_2
\ee
Similar to the case of bi-partition, we can decompose the bulk Hilbert space into: 
\be
\mathcal{H}_{code}=\oplus_{\alpha,\beta} \mathcal{H}_a(\alpha)\otimes \mathcal{H}_b(\alpha,\beta)\otimes \mathcal{H}_c(\beta) 
\ee 
where $\alpha$ and $\beta$ denote the induced metrics on $\Sigma_1$ and $\Sigma_2$ respectively, together with the conformal structures in the normal planes. Proceeding further we can decompose the bulk state into fixed-area states labeled by two numbers: 
\be\label{eq:decom_2areas}
|\psi\rangle_{\text{bulk}} = \sum_{\mathcal{A}_1,\mathcal{A}_2} \sqrt{P\left(\mathcal{A}_1,\mathcal{A}_2\right)}\;|\psi_{\mathcal{A}_1,\mathcal{A}_2} \rangle.
\ee
Although we have been summing fixed-area states over the areas, these sums can also be written as integrals.

Now let us continue the bulk computation of the Renyi entropy for the high energy black hole microstate, starting from Eq.~\eqref{zneff}:
\be\label{zneff2}
Z_n = \int \mathcal{D}g\mathcal{D}\phi \mathcal{D}\Sigma_A \mathcal{D}\Sigma_{\bar{A}}\;\; e^{-n I_{\text{bulk}}(g,\phi,E)-I_{\text{eff}}(\Sigma_A,\Sigma_{\bar{A}},n)}.
\ee
This path integral can be performed by postponing the integrals over the areas $\hat{\mathcal{A}}_A$, $\hat{\mathcal{A}}_{\bar{A}}$ of the cosmic branes $\Sigma_A$, $\Sigma_{\bar{A}}$ until the very end.  This amounts to first doing the other integrals in the fixed-area state $|\psi_{\mathcal{A},\bar{\mathcal{A}}}\rangle$ where the areas $\hat{\mathcal{A}}_A$, $\hat{\mathcal{A}}_{\bar{A}}$ are fixed to the values $\mathcal{A}$, $\bar{\mathcal{A}}$, and then doing the final integrals over the areas $\mathcal{A}$, $\bar{\mathcal{A}}$.  Therefore, we rewrite Eq.~\eqref{zneff2} as
\be\label{znp}
Z_n = \int d\mathcal{A} \, d\bar{\mathcal{A}} \,\, P_n(\mathcal{A},\bar{\mathcal{A}})  e^{-I_{\text{eff}}\left(\mathcal{A},\bar{\mathcal{A}},n\right)}
\ee
where we have defined
\be\label{pn}
P_n(\mathcal{A},\bar{\mathcal{A}}) \equiv \int \mathcal{D}g\mathcal{D}\phi\Big|_{\hat{\mathcal{A}}_A[g]=\mathcal{A},\, \hat{\mathcal{A}}_{\bar{A}}[g]=\bar{\mathcal{A}}} \; e^{-n I_{\text{bulk}}(g,\phi,E)}
\ee
and $I_{\text{eff}}\left(\mathcal{A},\bar{\mathcal{A}},n\right)$ is determined by evaluating Eq.~\eqref{eq:Sb_eff} in the fixed-area state:
\be\label{eq:Sb_eff2}
I_{\text{eff}}\left(\mathcal{A},\bar{\mathcal{A}},n\right)
= \begin{cases} 
\frac{n-1}{4 G_N}\mathcal{A} -\ln{\left(\mbox{$_2$F$_1$}\left[1-n,-n;\,2;\, e^{\frac{\Delta \mathcal{A}}{4 G_N}}\right]\right)},\; \Delta \mathcal{A}<0 \\
\frac{n-1}{4 G_N}\bar{\mathcal{A}} -\ln{\left(\mbox{$_2$F$_1$}\left[1-n,-n;\,2;\, e^{-\frac{\Delta \mathcal{A}}{4 G_N}}\right]\right)},\; \Delta \mathcal{A}>0
\end{cases}
\ee
with
\be
\Delta \mathcal{A} \equiv \mathcal{A} - \bar{\mathcal{A}}.
\ee

In the saddle point approximation, we can rewrite Eq.~\eqref{pn} and express it in terms of $P(\mathcal{A},\bar{\mathcal{A}})$ that we saw earlier:
\be\label{pnp}
P_n(\mathcal{A},\bar{\mathcal{A}}) = \mathcal{N}^n P(\mathcal{A},\bar{\mathcal{A}})^n
\ee
where
\be
P(\mathcal{A},\bar{\mathcal{A}}) = \mathcal{N}^{-1} \int \mathcal{D}g\mathcal{D}\phi\Big|_{\hat{\mathcal{A}}_A[g]=\mathcal{A},\, \hat{\mathcal{A}}_{\bar{A}}[g]=\bar{\mathcal{A}}} \; e^{-I_{\text{bulk}}(g,\phi,E)}.
\ee
Plugging Eq.~\eqref{pnp} into Eq.~\eqref{znp}, we find
\be\label{inp}
Z_n = \mathcal{N}^n \int d\mathcal{A} \, d\bar{\mathcal{A}} \,\, P(\mathcal{A},\bar{\mathcal{A}})^n  e^{-I_{\text{eff}}\left(\mathcal{A},\bar{\mathcal{A}},n\right)}.
\ee

For chaotic high energy eigenstates, $P(\mathcal{A},\bar{\mathcal{A}})$ is highly peaked on a codimension-one trajectory:\footnote{This peak becomes exactly a delta function in the limit that we are interested in.}
\be
 P(\mathcal{A},\bar{\mathcal{A}}) \approx P(\mathcal{A})\delta_{\bar{\mathcal{A}},\;\bar{\mathcal{A}}(\mathcal{A})}
\ee
where $\bar{\mathcal{A}}\left(\mathcal{A}\right)$ denotes a function of $\mathcal{A}$. More specifically, $\bar{\mathcal{A}}\left(\mathcal{A}\right)$ is given by the area of the extremal surface homologous to $\bar{A}$ evaluated in the saddle point geometry $g$ with fixed extremal area $\hat{\mathcal{A}}_A[g]=\mathcal{A}$, i.e., the saddle point solution for the Euclidean path integral
\be\label{eq:fixed-area_norm}
P(\mathcal{A})= \mathcal{N}^{-1} \int \mathcal{D}g\mathcal{D}\phi\Big|_{\hat{\mathcal{A}}_A[g]=\mathcal{A}} \; e^{-I_{\text{bulk}}(g,\phi,E)}.
\ee

For our interests, $|\psi\rangle_{\text{bulk}}$ is dual to a high temperature black hole. Correspondingly, the dominant saddle point solution $\mathcal{B}(\mathcal{A})$ for Eq. (\ref{eq:fixed-area_norm}) can be obtained by gluing two portions of black hole geometries $\mathcal{B}_A(E_1)$ and $\mathcal{B}_{\bar{A}}(E_2)$ along $A$ and $\bar{A}$, and whose ADM masses are $E_1$ and $E_2$ respectively, constrained by requiring that the total ADM mass of $\mathcal{B}$ is fixed to be $E$:
\be \label{eq:fixed-area_saddle}
\mathcal{B}(\mathcal{A}) = \mathcal{B}_A (E_1)\cup \mathcal{B}_{\bar{A}}(E_2),\; E_1 = f^{-1}\mathcal{E}(\mathcal{A}/f),\; E_2 = E-f E_1
\ee
where $\mathcal{E}(A)$ is the $f$-fractional ADM mass of a black hole with total horizon area $A$.  In $d+2$ dimensional bulk it is given by
\be\label{eq:area_mass}
\mathcal{E}(A)=\frac{f S^d\;d}{16\pi G_N L_{AdS}^2}\left(\frac{A}{S^d}\right)^{1+\frac{1}{d}},\;\;S^d = \frac{2\pi^{\frac{d+1}{2}}}{\Gamma\left(\frac{d+1}{2}\right)}.
\ee

Using the relation between horizon area and entropy, it is easy to derive
\be
 \mathcal{A} = 4G_N f V s\left[\frac{\mathcal{E}(\mathcal{A}/f)}{fV}\right],\;\bar{\mathcal{A}}\left(\mathcal{A}\right) = 4G_N (1-f)V s\left[\frac{E- \mathcal{E}\left(\mathcal{A}/f\right)}{(1-f)V}\right]
\ee
where $fV$ and $(1-f)V$ are the subsystem volumes $V_A$ and $V_{\bar{A}}$, and $s(e)$ is the boundary entropy density at energy density $e$.  At this step it is easy to switch the integration variable from $\mathcal{A}$ to $\mathcal{E}\left(\mathcal{A}/f\right)$ and rewrite Eq.~\eqref{inp} as
\bea
Z_n = \mathcal{N}^n \int d\mathcal{E} P(\mathcal{E})^n \, e^{-I_{\text{eff}}\left( f V s\left[\frac{\mathcal{E}}{fV}\right], (1-f)V s\left[\frac{E- \mathcal{E}}{(1-f)V}\right],n\right)},
\eea

To evaluate this integral, we identify the arguments in $I_{\text{eff}}$ with the subsystem entropies: 
\be
 f V s\left[\frac{\mathcal{E}}{fV}\right] = S_A(\mathcal{E}),\;(1-f)V s\left[\frac{E-\mathcal{E}}{(1-f)V}\right] = S_{\bar{A}}(E-\mathcal{E}).
\ee
We also rewrite Eq.~(\ref{eq:fixed-area_saddle}) as
\be
\mathcal{N} P(\mathcal{E}) = \int \mathcal{D}g\mathcal{D}\phi\Big|_{\hat{\mathcal{A}}_A[g]=f A\left(\mathcal{E}\right)} \; e^{-I_{\text{bulk}}(g,\phi,E)} 
\ee
where $A\left(\mathcal{E}\right)$ is the inverse of $\mathcal{E}(A)$, and in particular $fA\left(\mathcal{E}\right) = S_A(\mathcal{E})$.  In the saddle point approximation, we find simply
\be
\mathcal{N}P(\mathcal{E}) \approx e^{S_A\left(\mathcal{E}\right)+S_{\bar{A}}(E-\mathcal{E})}.
\ee 
Putting these together, we arrive at
\be 
Z_n =  \int d\mathcal{E}\; e^{nS_A\left(\mathcal{E}\right)+nS_{\bar{A}}(E-\mathcal{E})-I_{\text{eff}}\left( S_A(\mathcal{E}), S_{\bar{A}}(E-\mathcal{E}),n\right)}.
\ee
Plugging the form of $I_{\text{eff}}$ from Eq.~(\ref{eq:Sb_eff2}) into this equation, we find that it exactly reproduces Eq.~(\ref{eq:trace_intergal}) which we derived from the chaotic ansatz and already calculated in Sec.~\ref{sec:general}.  In other words, we have re-derived Eq.~(\ref{eq:trace_intergal}) using a holographic calculation. The enhanced correction proportional to $\sqrt{V}$ near the transition $f=1/2$ therefore follows from the analysis in Sec.~\ref{sec:general}. 

Let us quickly summarize the main steps in this section. We have shown that for black hole microstates, averaging over randomness gives rise to different gluing boundary conditions $\lbrace\mathcal{M}_m\rbrace$ for computing $\text{tr}\overline{\rho_A^n}$, which are filled by different bulk saddle point geometries $\lbrace \mathcal{B}_m \rbrace$, most of which break replica-symmetry. However, we argued that when working in the high energy limit and neglecting boundary effects, an approximate $U(1)$ rotation-symmetry emerges for each saddle. Utilizing this symmetry then allows us to construct the quotient geometry of each saddle $\tilde{\mathcal{B}}_m = \mathcal{B}_m/\mathbb{Z}_n$, where $\mathbb{Z}_n \subset U(1)$, using the \textit{double-defect} construction, i.e. inserting two segments of cosmic branes covering $A$ and $\bar{A}$ and with brane-tensions $(T_A, T_{\bar{A}})$ respectively. The double-defect insertions for distinct saddles can be re-summed into a (non-local) effective action $I_{\text{eff}}$ for the cosmic branes. The effective action can be evaluated in the fixed-area basis, which for the black hole microstates is conveniently associated with the subsystem energy eigenstates. Performing this calculation then reproduces exactly our results in Sec.~\ref{sec:general}.

\section{Discussion}\label{sec:disc}

In this paper, we studied the details of entanglement transition in subsystem Renyi entropy $S^A_n$ in the context of high energy eigenstates. Our analysis started with the chaotic ansatz (\ref{eq:random_z}), and is later repeated for the high energy black hole states in AdS/CFT. We focused on the $\mathcal{O}\left(\sqrt{V}\right)$ enhanced correction to the microstate Renyi entropy near the transition as we tune the sub-system size $V_A = f V$ towards half-point $f=1/2$. We found that such enhanced corrections emerge in the Renyi entropy when the Renyi index $n$ is sufficiently close to 1, i.e., if $n-1 \ll 1/\sqrt{C_V}$. It is consistent and connects with previous results for von Neumann entropy in \cite{Srednicki:2019}. Below, we remark and discuss a few points regarding our analysis, as well as general lessons our computation seems to be suggesting. 

Let us revisit the proposed chaotic ansatz (\ref{eq:random_z}), especially the nature of the random variables $c_{iJ}$. We implicitly assumed that they are independent Gaussian variables. This assumption shows up (see Appendix~\ref{app:contraction}) when computing the disorder average, in the form of only including contributions from Wick contractions $\overbracket{c c}$. There is no ``interaction'' contributions, e.g. $\overbracket{ccc}$. Technically, this made the calculation tractable and one could obtain explicit result, e.g. $G_n$ in (\ref{eq:renyi_wick}), that allows for analytic continuation in $n$. Being independent Gaussian variables is a fairly strong assumption on $c_{iJ}$. However, the analysis in Sec.~\ref{sec:general} for the enhanced correction is highly independent of the specific functional form of $G_n$, as long as it smoothly connects the two colliding saddle points in the limit $n\to 1$, thus providing an approximately flat interval. This suggests that the enhanced correction is insensitive to the precise random distribution of $c_{iJ}$. In AdS/CFT, the independent Gaussian distribution for the randomness $\hat{c}$ manifests in the type of boundary manifolds $\mathcal{M}_m$ that emerges after averaging, in particular the gluing options for $\mathcal{M}_m$. Only pairwise gluing between upper and lower boundaries of the Euclidean path integral for $\rho$ is included. ``Interaction'' among the $\hat{c}$ 's would result in gluing using a ``multi-way junction''. It is likely that in holography, bulk saddles resulting from such gluing is suppressed in $G_N$, making the independent Gaussian distribution natural at the leading order in $G_N$. We leave these for future investigations. 

An important part in our analysis is the treatment of ``intermediate'' saddles that do not show replica-symmetry explicitly. In general they are very difficult to study, especially in holography. As a result these saddles are usually ignored under the belief that even if exist, they do not play important roles. However, such attitude becomes questionable near entanglement transitions. In our case, we were able to construct and calculate these replica-nonsymmetric saddles $\mathcal{B}_m$ explicitly thanks to the emergent $U(1)$ rotation-symmetry in the high energy limit. This gave us the opportunity to examine how they participate in the relevant physics, i.e. enhanced correction to the microstate Renyi entropy, close to the transition. Recall that in Sec.~\ref{sec:general} we have shown that the enhanced correction can be entirely captured by simply including only the dominant replica-symmetric configuration: 
\be
\int d\mathcal{E}\; e^{F_1(\mathcal{E})} \approx \int d\mathcal{E}\;e^{F_{\text{dom}}\left(\mathcal{E}\right)}
\ee
where
\be
e^{F_{\text{dom}}\left(\mathcal{E}\right)} = e^{S_A\left(\mathcal{E}\right)+S_{\bar{A}}\left(E-\mathcal{E}\right)}\text{max}\left\lbrace e^{(n-1)S_A(\mathcal{E})}, e^{(n-1)S_{\bar{A}}(E-\mathcal{E})}\right\rbrace.
\ee
There we observed that the remaining saddle point configurations only give up to $\mathcal{O}\left(1\right)$ correction to $S^{\text{dom}}_n$. This supports the approach of only keeping the dominant replica-symmetric saddle point contribution, taken in Ref.~\cite{Srednicki:2019} to find the enhanced correction. On the other hand, one may hope to ``dynamically'' implement the prescription of picking the dominant replica-symmetric saddle. A natural guess might be simply adding up contributions from both replica-symmetric saddle points, and consider: 
\bea
e^{F_{\text{sym}}\left(\mathcal{E}\right)} &=& e^{S_A\left(\mathcal{E}\right)+S_{\bar{A}}\left(E-\mathcal{E}\right)} \left[e^{(n-1)S_{\bar{A}}\left(E-\mathcal{E}\right)}+e^{(n-1)S_A\left(\mathcal{E}\right)}\right]\nonumber\\
&=& \exp{\left[F_{\text{dom}}\left(\mathcal{E}\right)+ F^{\text{sym}}_\Delta\left(\mathcal{E}\right)\right]}\nonumber\\
F^{\text{sym}}_\Delta\left(\mathcal{E}\right) &=& \ln{\left(1+ e^{-(n-1)|S_A\left(\mathcal{E}\right)-S_{\bar{A}}\left(E-\mathcal{E}\right) |}\right)}
\eea
where we have decomposed it into a dominant-saddle part $F_{\text{dom}}$ and a correction part $F^{\text{sym}}_\Delta$, similar to the case of $F_1$.  Intuitively one may expect $F_{\text{sym}}$ to be a good ``proxy'' for $F_{\text{dom}}$ and thus also for $F_1$. However, one can immediate see that this is not the case. Due to the reflection symmetry at $f=1/2$, we have that:
\be 
\int d\mathcal{E}\;e^{F_{\text{sym}}\left(\mathcal{E}\right)} \equiv 2\times \int d\mathcal{E}\;e^{F_2\left(\mathcal{E}\right)} 
\ee 
Therefore the correction $S^{\text{sym}}_n-S^{\text{MC}}_n \equiv \frac{\ln 2}{1-n}$, independent of the total volume $V$ for all $n$. In terms of separating into the dominant part and correction part, $F^{\text{sym}}_\Delta$ is still bounded by $V$-independent constants: 
\be
0\leq F^{\text{sym}}_\Delta\left(\mathcal{E}\right)\leq \ln{2}
\ee 
However, the bound does not shrink towards $F^{\text{sym}}_\Delta \approx 0$ as $n\to 1$, which is what happened for $F_\Delta$ of $F_1$, and hence the difference between $S^{\text{sym}}_n$ and $S^{\text{dom}}_n$ becomes large. In this sense, $F_{\text{sym}}$ is no longer a good proxy to both $F_{\text{dom}}$ and $F_1$ in the regime of enhancement. In summary, it suggests that to capture the enhanced correction near transition, it is sufficient to include only the dominant replica-symmetric saddle, which gives $S^{\text{dom}}_n$; however, we would miss such effect by simply adding up the two replica-symmetric saddles and hoping it ``dynamically'' mimics $S^{\text{dom}}_n$. Re-summing over the replica-nonsymmetric saddles is important for the purpose of implementing the prescription of $S^{\text{dom}}_n$. 

Our holographic calculation in Sec.~\ref{sec:holography} for black hole microstates have a similar feature: instead of summing over all fixed-area saddle points, we can simply choose to keep only the dominant replica-symmetric one, but we should not try to sum over both replica-symmetric saddle points.  Concretely speaking, the replicated partition function $Z_n$ for calculating Renyi entropies is given by Eq.~\eqref{inp} and involves an effective action $I_{\text{eff}}$ that includes the contributions from all fixed-area saddle points, but we could replace it by the dominant replica-symmetric contribution and write Eq.~\eqref{inp} approximately up to $\mathcal{O}(1)$ errors as
\be
Z_n = \mathcal{N}^n \int d\mathcal{A} \, d\bar{\mathcal{A}} \,\, P(\mathcal{A},\bar{\mathcal{A}})^n  e^{-I_{\text{eff}}\left(\mathcal{A},\bar{\mathcal{A}},n\right)}
\approx \mathcal{N}^n \int d\mathcal{A} \, d\bar{\mathcal{A}} \,\, P(\mathcal{A},\bar{\mathcal{A}})^n  e^{-I_{\text{dom}}\left(\mathcal{A},\bar{\mathcal{A}},n\right)}
\ee
where instead of Eq.~\eqref{eq:Sb_eff2} we have used only the on-shell action of the dominant fixed-area saddle
\be \label{eq:Sbdom}
I_{\text{dom}} \left(\mathcal{A},\bar{\mathcal{A}},n\right) = \frac{n-1}{4 G_N} \min \left\{ \mathcal{A},\bar{\mathcal{A}} \right\}.
\ee
From this, we can take the $n\to 1$ limit and find the von Neumann entropy
\be\label{eq:vN_prescription}
S^A_{\text{ent}} \approx \int d\mathcal{A} \, d\bar{\mathcal{A}} \,\, P(\mathcal{A},\bar{\mathcal{A}}) \frac{\min \left\{ \mathcal{A},\bar{\mathcal{A}} \right\}}{4 G_N}.
\ee

\begin{figure}[h!]
\centering
\includegraphics[scale=0.4]{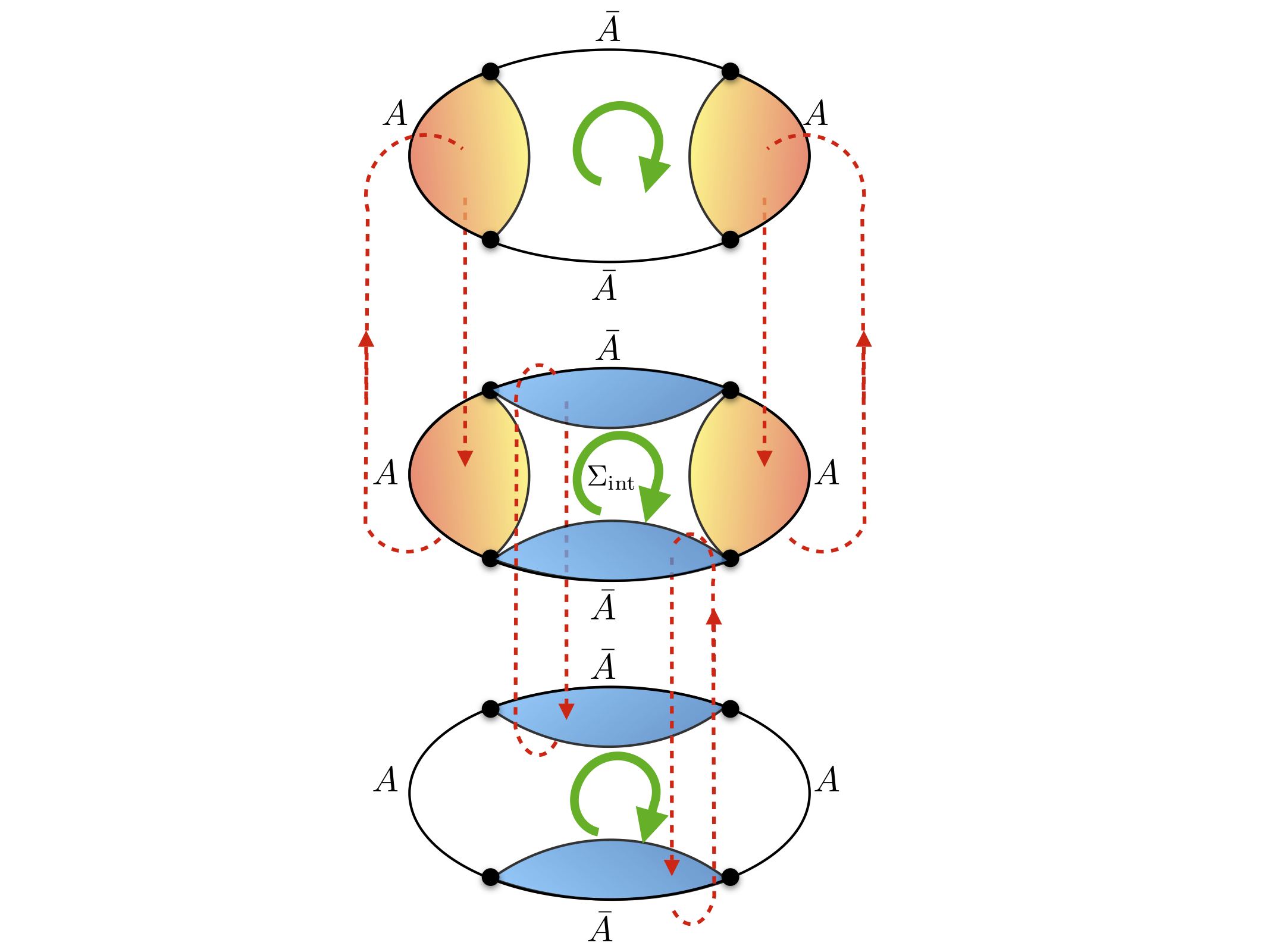}  
\caption{ In the case of two-interval subsystem $A$ in $\text{AdS}_3/\text{CFT}_2$, the topology for a replica-nonsymmetric saddle for $n=3$. 
}
\label{fig:2_interval}
\end{figure}

It would be very interesting to understand to what extent our results for black hole microstates apply to more general holographic states with HRT entanglement transitions, such as the simple example of two intervals in the vacuum state of $\text{AdS}_3/\text{CFT}_2$.  In these general examples, if we can similarly replace the effective action $I_{\text{eff}}$ from summing over all fixed-area saddle points with the dominant replica-symmetric contribution $I_{\text{dom}}$, we would by the same steps find enhanced corrections to the holographic entanglement entropy from evaluating Eq.~\eqref{eq:vN_prescription}.  For our example of the high energy black hole microstates, we argued that such a replacement is a legitimate approximation, by computing explicitly the correction due to re-summing other saddles and showing it remains $\mathcal{O}(1)$ near transition as $n\to 1$. We can ask if similar arguments can be made for more general holographic examples such as the two-interval subsystem, by re-summing and analyzing the contributions from replica-nonsymmetric saddles\footnote{One detailed difference is that in a typical holographic example such as the two-interval case, the bulk saddles being summed over all have the same replica-symmetric boundary, whereas in our black hole example, the bulk saddles $\mathcal{B}_i$ have different boundaries $\mathcal{M}_i$ each of which might not respect the replica symmetry.}. In terms of topology, the gluing prescription for constructing these saddles can proceed in ways that are analogous to the high energy black hole case (see Figure~\ref{fig:2_interval}). However, due to the additional region $\Sigma_{\text{int}}$ between the two extremal surfaces, we can no longer take the replica-symmetric quotient for each of the saddles as we did for the black hole microstates\footnote{Such regions $\Sigma_{\text{int}}$ are absent or negligible in high energy black hole geometries (see Figure~\ref{fig:bulk}).}. One might instead consider the two-interval subsystem in an excited global state, whose bulk geometry is deformed such that $\Sigma_{\text{int}}$ shrinks to being negligible, i.e. the two competing extremal surfaces almost touch each other. However, the emergent $U(1)$ rotation-symmetry we had for the black hole states is still missing, as the symmetry-breaking features resulting from gluing will in this case affect the bulk defects extensively (though it is possible that a discrete $\mathbb{Z}_n$ symmetry may survive, we leave this possibility for future considerations).  In any case, it is difficult to re-sum their contributions into an explicit effective action $I_{\text{eff}}$ in terms of cosmic brane defects. For this reason we cannot make a concrete argument for neglecting the corrections from the replica-nonsymmetric saddles to the dominant one near transitions. As remarked before, the explicit form of $I_{\text{eff}}$ is not crucial for our analysis of the enhanced correction, as long as we include all saddles and not just the replica-symmetric ones. Hopefully the lack of explicit results can be accommodated by such flexibility, and the validity of neglecting these corrections even near transitions can be extended to more generic cases in AdS/CFT. 

We end this paper with a few future directions. Firstly, for the black hole microstates calculation, the emergence of additional saddles can be understood as coming from averaging over the randomness. Certain chaotic effects will manifest semi-classically in Euclidean path-integral after this, as is usually the case. It would be interesting to investigate how the same effects manifest in individual state without averaging. The emergence of these saddles break the topological obstruction for the RT surface due to the event horizon, understanding them for individual states would therefore shed light on probing the interior of black holes. Secondly, in this paper we solved the effective action $I_{\text{eff}}$ for cosmic branes in the fixed-area basis, which then reduced the calculations to the analysis in Sec.~\ref{sec:general}. It would be interesting to solve $I_{\text{eff}}$, or at least extract the relevant physics in $I_{\text{eff}}$ responsible for the enhanced corrections, in terms of possibly non-perturbative dynamics between the cosmic brane defects. By doing this, the cosmic brane defects may reveal themselves as objects with intrinsic dynamics, this may lead to new perspectives in entanglement calculations. Lastly, it would be very interesting and important to make concrete arguments for the order of corrections the replica-nonsymmetric saddles produce near transitions. We can phrase this problem in terms of properties of a correction term similar to $F_\Delta$ in Sec.~\ref{sec:general}. No doubt this will be very difficult, especially to extract its general properties and at the same time analytically continue the Renyi index $n\to 1$. One possible route is to apply the resolvent trick as in \cite{replica_wormhole_west} and translate properties one can infer at integer $n$ to general values of $n$. We plan to pursue such investigations in future works. 

\acknowledgments
We thank Don Marolf for many discussions related to this work.  It is also a pleasure to thank Chris Akers, Geoffrey Penington, Xiaoliang Qi, Pratik Rath, Mark Srednicki, and Douglas Stanford for useful conversations.  This material is based upon work supported by the Air Force Office of Scientific Research under award number FA9550-19-1-0360.  XD is also supported in part by funds from the University of California.  HW is supported by the Gordon and Betty Moore Foundation through Grant GBMF7392 at the Kavli Institute for Theoretical Physics (KITP).  This work was developed in part at the KITP which is supported in part by the National Science Foundation under Grant No.\ PHY-1748958.

\appendix
\section{Disorder Wick contraction}\label{app:contraction}
In computing the microstate Renyi entropy disorder-averaged, one needs to evaluate the following sum: 
\bea\label{eq:wick_contraction}
\overline{\sum_{a_1,...,a_n;b_1,...,b_n} c_{a_1,b_1}z_{a_2,b_1}z_{a_2,b_2}c_{a_3,b_2},...,c_{a_n,b_n}z_{a_1,b_n}}= \sum \left(\text{all Wick contractions}\right).
\eea
Let us examine in more details the structure of contractions. We can label each random variable $c_{a,b}$ by a circular dot with two legs extending out, with a lower leg representing the $a$ index and an upper leg representing the $b$ index. The summation $\sum_{a_1,...,a_n;b_1,...,b_n}$ suggests to pairwise connect these indices. For illustration we connect the $a$ indices using red lines, and $b$ indices using green lines. The contraction proceeds by
\be
\overbracket{c_{a,b}c_{a',b'}}=\delta_{a,a'}\delta_{b,b'} 
\ee
\begin{figure}[h!]
\centering
\includegraphics[scale=0.32]{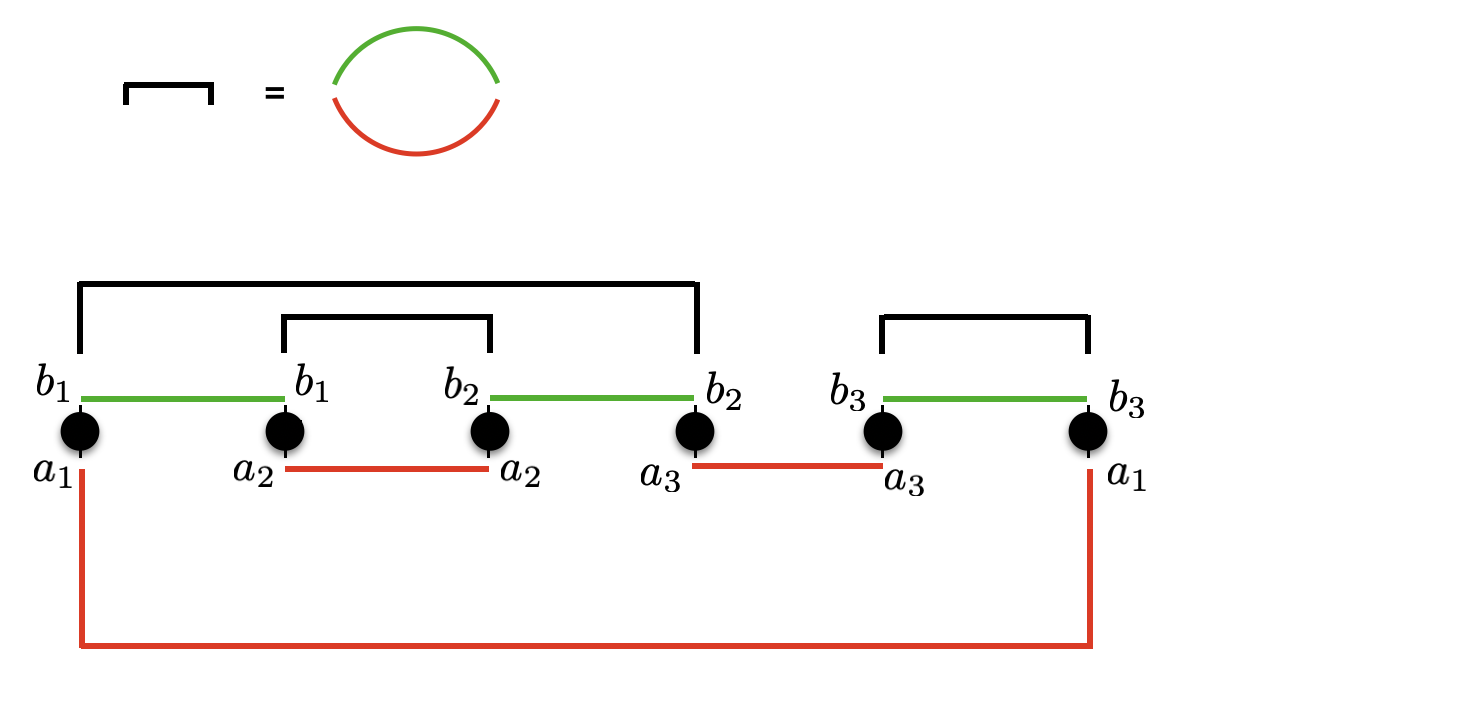}  
\includegraphics[scale=0.32]{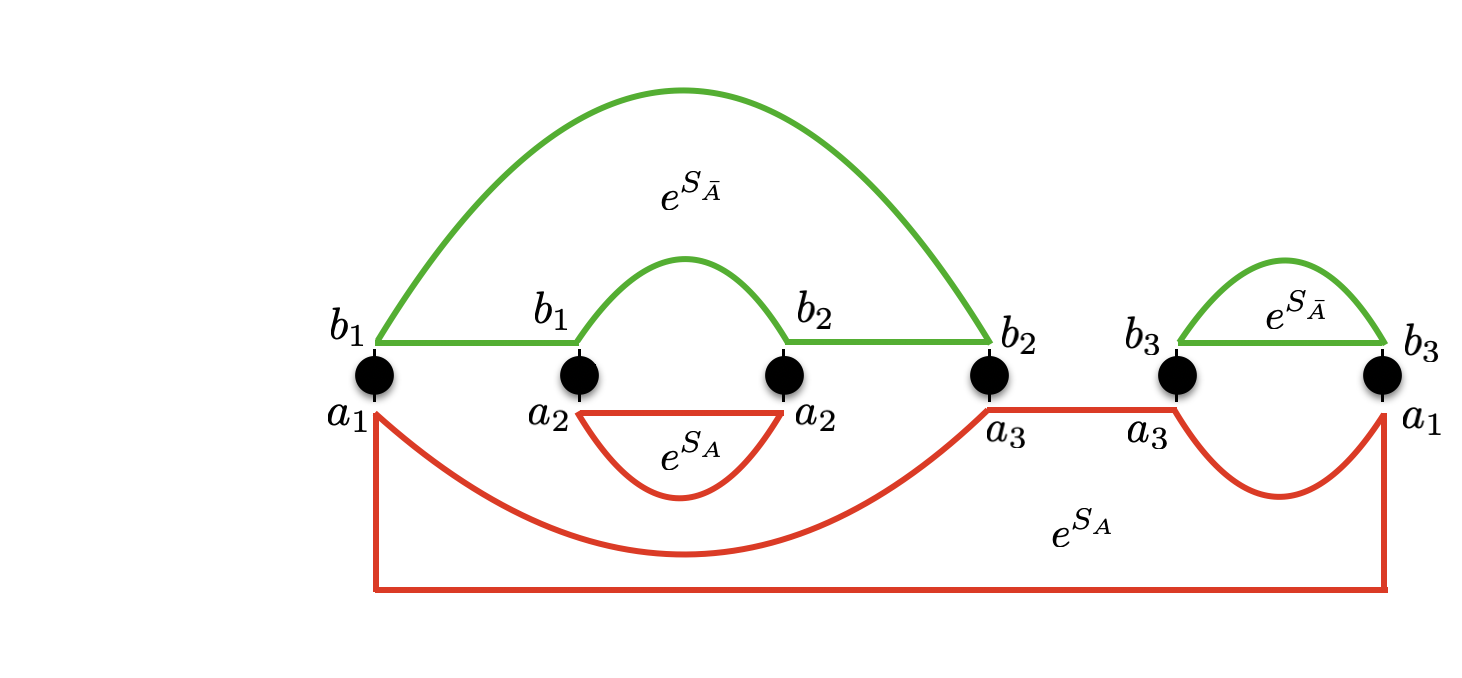}  
\caption{An example of contractions (left) producing loop diagrams (right) for the case of $n=3$
}
\label{fig:contraction loop}
\end{figure}
We can therefore represent each contraction $\overbracket{c_{a,b}c_{a',b'}}$ graphically by a pair red and green lines connecting $(a,a')$ and $(b,b')$ respectively. They then combine with the red and green lines coming from $\sum_{a_1,...,a_n;b_1,...,b_n}$ to form closed loops  (see Figure~\ref{fig:contraction loop}). Each red loop contributes a factor of $e^{S_A}$;  each green loop contributes a factor of $e^{S_{\bar{A}}}$. where $e^{S_A}, e^{S_{\bar{A}}}$ are the rank of vector spaces associated with the indices $a$ and $b$ respectively. From the standard loop counting, we can conclude that the dominant contributions come from ``planar'' contractions, i.e. those that produce planar loop diagrams. It is easy to see that the planar contractions are characterized by non-intersection. For example, the following is a planar contraction: 
\be 
\overbracket{c_{a_1,b_1}\overbracket{c_{a_2,b_1}c_{a_2,b_2}}c_{a_3,b_2}}...\overbracket{c_{a_n,b_n}c_{a_1,b_n}} \sim e^{2 S_A} e^{(n-1) S_{\bar{A}}}
\ee
while the following is a non-planar contraction: 
\be 
 \rlap{$\overbracket{\phantom{c_{a_1,b_1}c_{a_2,b_1}c_{a_2,b_2}}}$}c_{a_1,b_1} \underbracket{c_{a_2,b_1}c_{a_2,b_2}c_{a_3,b_2}}...\overbracket{c_{a_n,b_n}c_{a_1,b_n}}\sim e^{S_A} e^{(n-1)S_{\bar{A}}}
\ee

\begin{figure}[h!]
\centering
\includegraphics[scale=0.17]{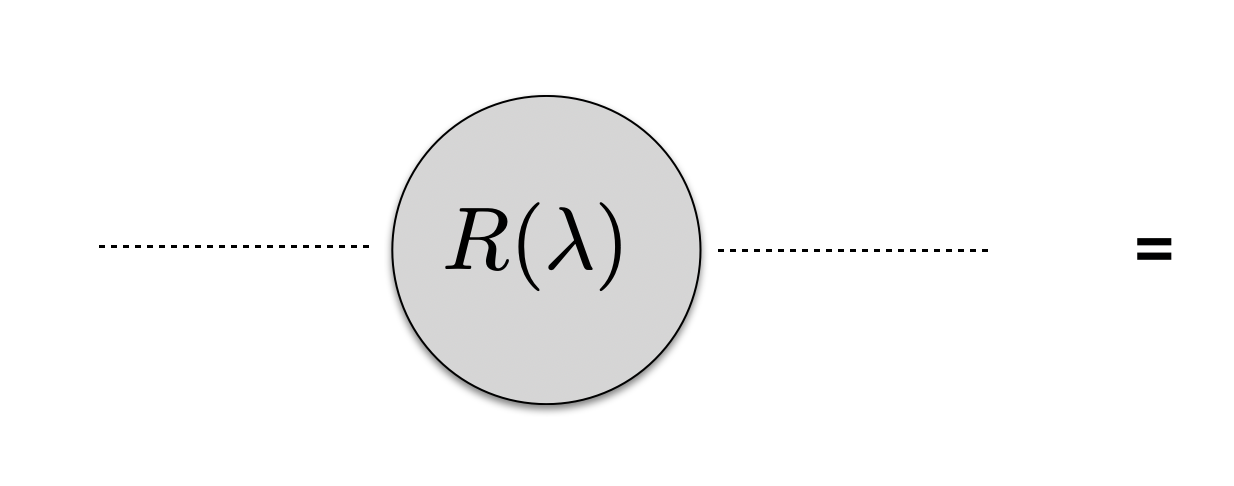}  
\includegraphics[scale=0.17]{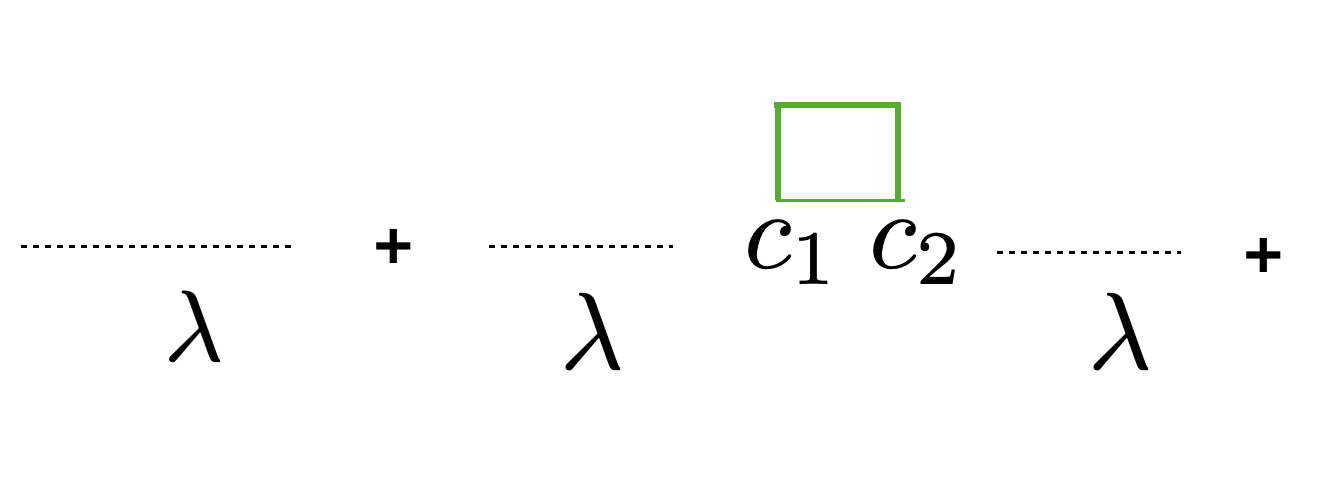}  
\includegraphics[scale=0.17]{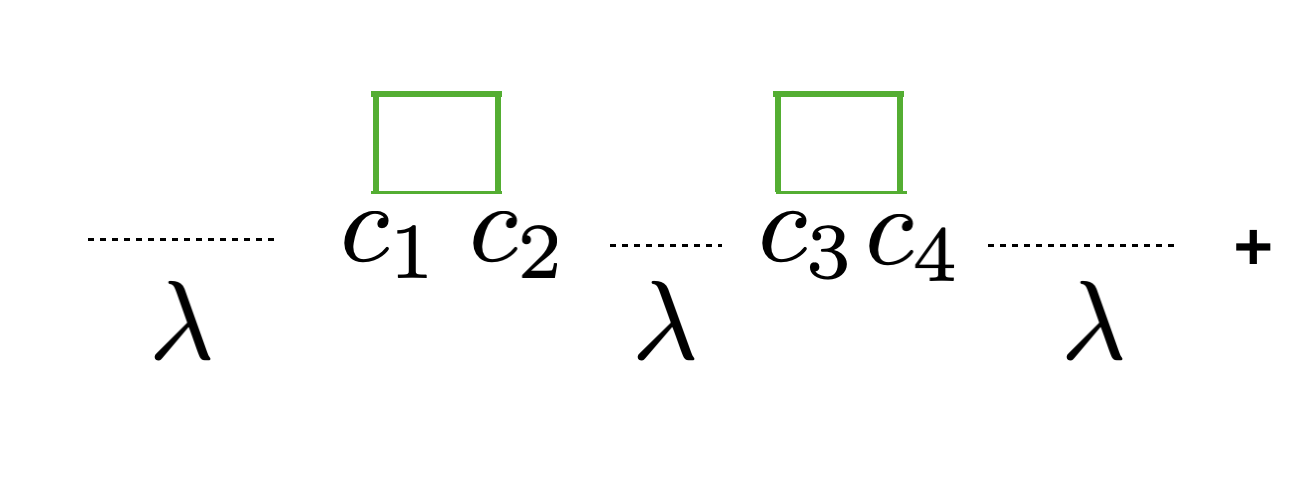}
\includegraphics[scale=0.17]{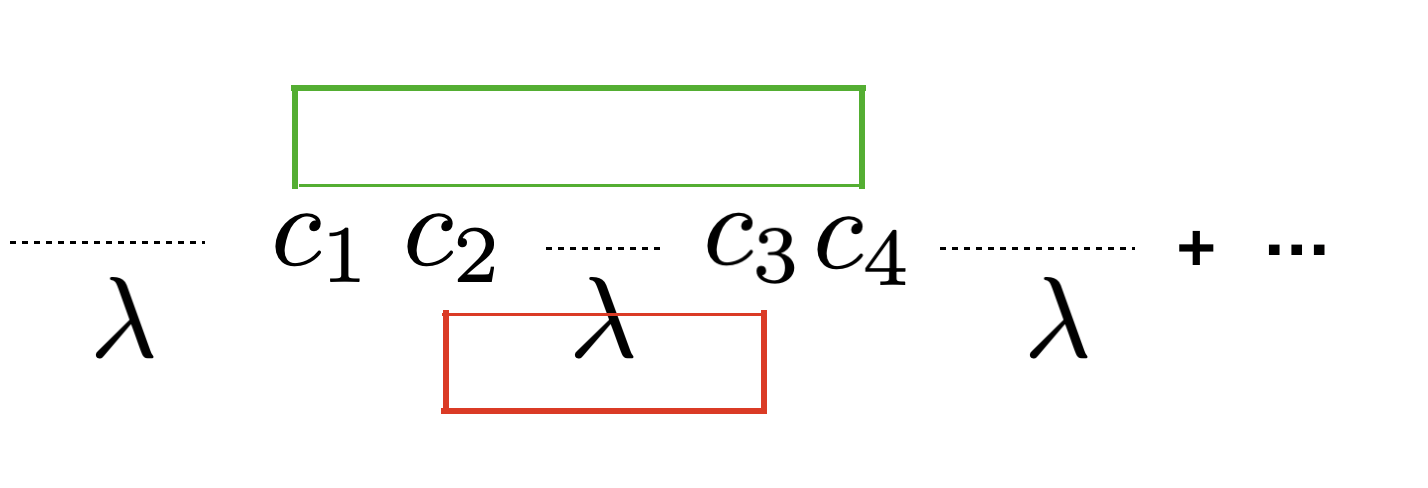}    
\includegraphics[scale=0.17]{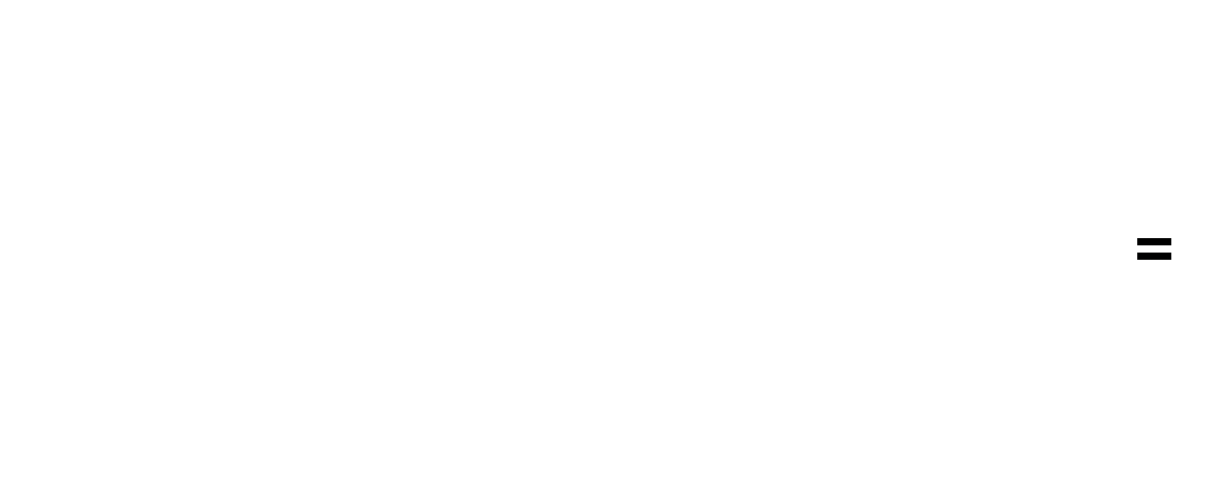}  
\includegraphics[scale=0.17]{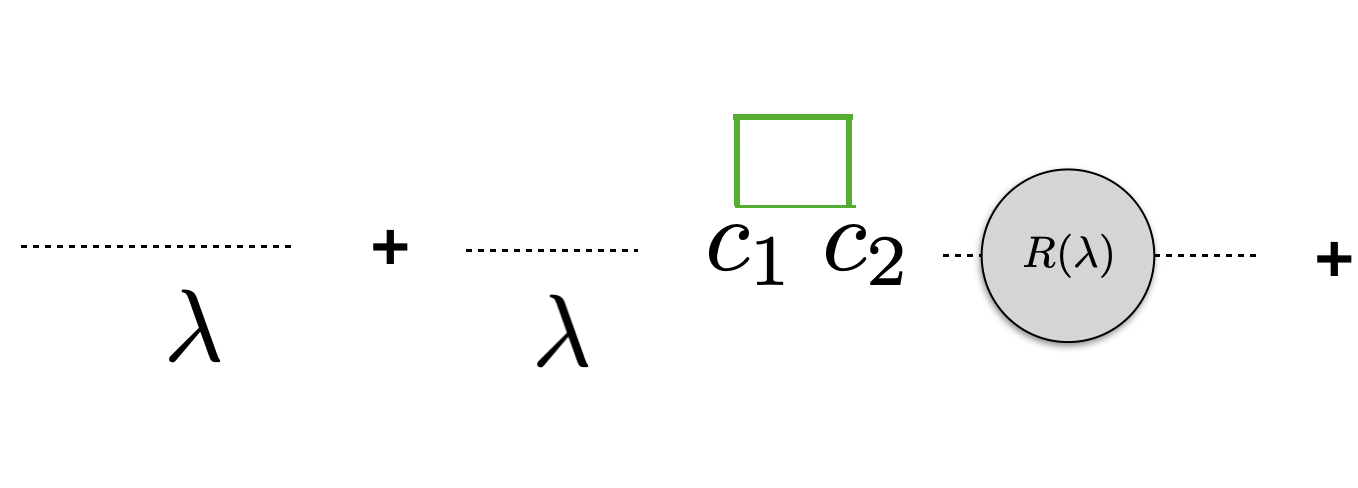} 
\includegraphics[scale=0.17]{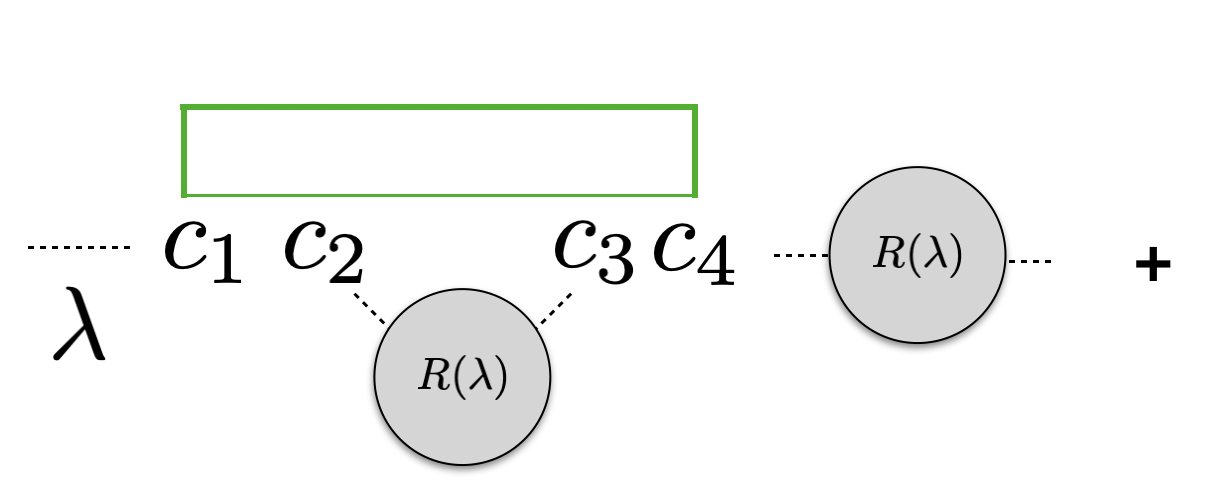}
\includegraphics[scale=0.17]{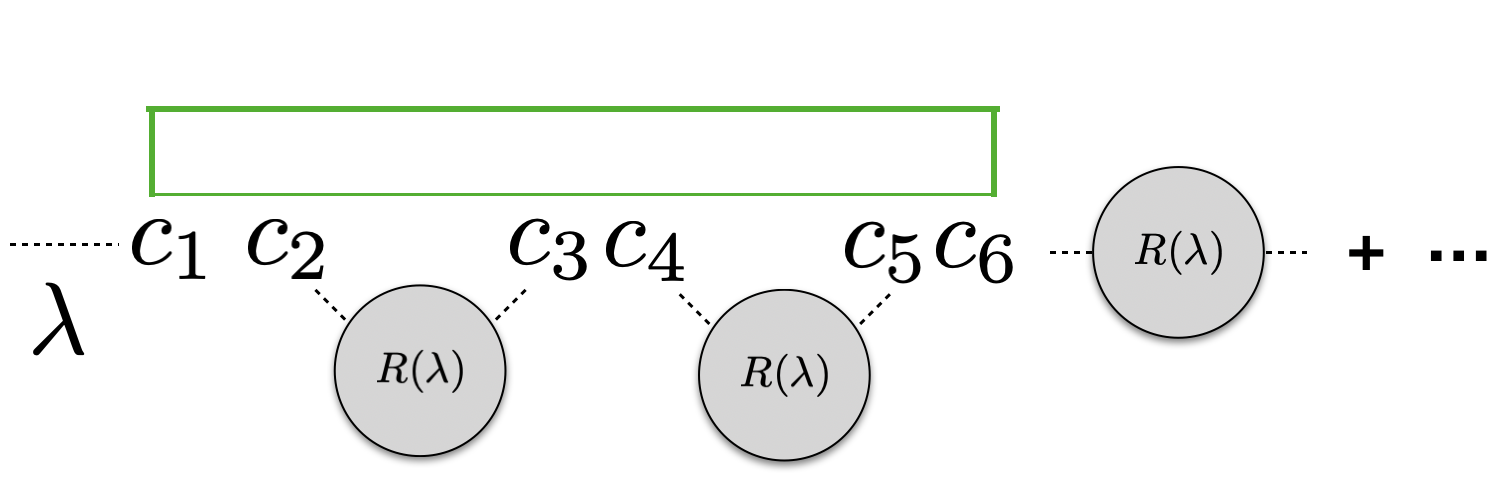}      
\caption{Schwinger-Dyson equation for the resolvent $R(\lambda)_{ij}$ from summing planar contractions.
}
\label{fig:resolvent_eq}
\end{figure}

In order to sum over all planar contractions, we need to work out the details relating contraction patterns and the loop factors, and it is a very complicated combinatorics exercise to directly do this. One way to get around this and proceed is to follow \cite{replica_wormhole_west} and consider the resolvent operator, which is defined by: 
\be\label{eq:res_def} 
R(\lambda)_{ij} = \left(\frac{\lambda}{1-\lambda \rho_A}\right)_{ij} = \lambda \delta_{ij} + \sum^{\infty}_{n=1} \lambda^{n+1} \left(\rho^n_A\right)_{ij}
\ee
We can focus on the disorder averaged version of the resolvent: 
\be 
\overline{R(\lambda)}_{ij} = \lambda \delta_{ij} + \sum^{\infty}_{n=1} \lambda^{n+1} \left(\overline{\rho^n_A}\right)_{ij}
\ee
From now on we simply write $\overline{R}(\lambda)$ as $R(\lambda)$. It is given by the sum over all planar contractions for all $n$ dressed by factors of $\lambda$, with the first and last index $(a_1=i, a_{n+1}=j)$ left un-contracted in each term; see top of Figure~\ref{fig:resolvent_eq}. Due to the planar nature of the contractions, there is a recursive structure that allows one to write down a self-consistent Schwinger-Dyson equation, graphically represented by the bottom of Figure~\ref{fig:resolvent_eq}: 
\bea 
R(\lambda)_{ij} = \lambda \delta_{ij} + \lambda e^{S_{\bar{A}}}\sum^{\infty}_{n=0} R(\lambda)^{n} R(\lambda)_{ij} =  \lambda \delta_{ij} +\frac{\lambda e^{S_{\bar{A}}}R(\lambda)_{ij}}{1-R(\lambda)}
\eea 
where $R(\lambda) = R(\lambda)_{ii}$ is the trace of the resolvent. 

Taking the trace of the Schwinger-Dyson equation gives us: 
\be 
R(\lambda) = \lambda e^{S_A} + \frac{\lambda e^{S_{\bar{A}}}R(\lambda)}{1-R(\lambda)}
\ee
whose solution is simply\footnote{There is another branch of the solution but its expansion is not inconsistent with Eq.~(\ref{eq:res_def}).}
\be
R(\lambda) = \frac{1+\left(e^{S_A}-e^{S_{\bar{A}}}\right)\lambda -\sqrt{\left[1+\left(e^{S_A}-e^{S_{\bar{A}}}\right)\lambda\right]^2-4 e^{S_A}\lambda}}{2} 
\ee
With a little re-packaging, we can write it as
\bea 
R(\lambda) &=& \lambda e^{S_{\bar{A}}}\; G\left(\lambda e^{S_A}, e^{S_{\bar{A}}-S_A}\right) + \lambda e^{S_A}\nonumber\\
G(z,t)&=&\frac{1-z(t+1)-\sqrt{1-2z(t+1)+z^2(t-1)^2}}{2tz} 
\eea
$G(z,t)$ is the generating function for the so-called Narayana numbers $N(n,m)$:
\be 
G(z,t) = \sum^\infty_{n=1}\sum^n_{k=1}N(n,k) z^n t^{k-1}
\ee
In combinatorics, $N(n,k)$ is equal to the number of planar contractions among $n$ pairs that contain $k$ distinct nestings, and is given by: 
\be 
N(n,m)=\frac{1}{n}\begin{pmatrix}
n\\
m
\end{pmatrix}\begin{pmatrix}
n\\
m-1
\end{pmatrix}
\ee
We can compare the expansions:
\bea 
R(\lambda) &=& \lambda e^{S_A} + \sum^{\infty}_{n=1} \lambda^{n+1} \text{tr}\left(\overline{\rho^n_A}\right)\nonumber\\
&=& \lambda e^{S_A} + \lambda e^{S_{\bar{A}}}\sum^\infty_{n=1}\sum^n_{k=1}N(n,k) \left(\lambda e^{S_A}\right)^n e^{(k-1)\left(S_{\bar{A}}-S_A\right)}
\eea
and extract the following closed-form expression: 
\bea
 \text{tr}\left(\overline{\rho^n_A}\right) &=& e^{S_{\bar{A}}+n S_A}\sum^n_{k=1}N(n,k) e^{(k-1)\left(S_{\bar{A}}-S_A\right)}\nonumber\\
&=& \begin{cases} e^{S_A} e^{n S_{\bar{A}}}\; \mbox{$_2$F$_1$}\left(1-n,-n;\,2;\,e^{S_A-S_{\bar{A}}}\right), \;S_A < S_{\bar{A}} \\e^{S_{\bar{A}}} e^{n S_A}\; \mbox{$_2$F$_1$}\left(1-n,-n;\,2;\,e^{S_{\bar{A}}-S_A}\right), \;S_A > S_{\bar{A}}
\end{cases}
\eea
This result is symmetric upon switching $S_A \leftrightarrow S_{\bar{A}}$. When $n$ is an integer, the two piece-wise branches coincide for all $\lbrace S_A, S_{\bar{A}}\rbrace $.

\bibliographystyle{JHEP}
\bibliography{ETH}
\end{document}